\documentclass[letterpaper]{JHEP3} 
\usepackage{epsfig} 
\pdfoutput=1 
\usepackage{accents} 
\usepackage{nicefrac} 
\usepackage{feynmf} 

\def \lyell#1{\pdfsetcolor{0 0 1 0}{#1}\pdfsetcolor{0 0 0 1}}
\def \lgree#1{\pdfsetcolor{.65 0 .65 0}{#1}\pdfsetcolor{0 0 0 1}}

\def \lblue#1{\pdfsetcolor{.6 .6 0 0}{#1}\pdfsetcolor{0 0 0 1}}

\def \lred#1{\pdfsetcolor{0 .65 .65 0}{#1}\pdfsetcolor{0 0 0 1}}

\def \white#1{\pdfsetcolor{0 0 0 0}{#1}\pdfsetcolor{0 0 0 1}}
\def \lgray#1{\pdfsetcolor{0 0 0 .15}{#1}\pdfsetcolor{0 0 0 1}}
\def \mgray#1{\pdfsetcolor{0 0 0 .4}{#1}\pdfsetcolor{0 0 0 1}}
\def \dgray#1{\pdfsetcolor{0 0 0 .7}{#1}\pdfsetcolor{0 0 0 1}}
\def \myell#1{\pdfsetcolor{0 0 1 .05}{#1}\pdfsetcolor{0 0 0 1}}
\def \dyell#1{\pdfsetcolor{0 0 1 .3}{#1}\pdfsetcolor{0 0 0 1}}
\def \mrora#1{\pdfsetcolor{0 .5 1 .03}{#1}\pdfsetcolor{0 0 0 1}}
\def \drora#1{\pdfsetcolor{0 .5 1 .25}{#1}\pdfsetcolor{0 0 0 1}}
\def \mred#1{\pdfsetcolor{0 1 1 .15}{#1}\pdfsetcolor{0 0 0 1}}
\def \dred#1{\pdfsetcolor{0 1 1 .4}{#1}\pdfsetcolor{0 0 0 1}}
\def \mygre#1{\pdfsetcolor{.2 0 1 .25}{#1}\pdfsetcolor{0 0 0 1}}
\def \dygre#1{\pdfsetcolor{.2 0 1 .5}{#1}\pdfsetcolor{0 0 0 1}}
\def \mgree#1{\pdfsetcolor{1 0 1 .25}{#1}\pdfsetcolor{0 0 0 1}}
\def \dgree#1{\pdfsetcolor{1 0 1 .4}{#1}\pdfsetcolor{0 0 0 1}}
\def \mbvio#1{\pdfsetcolor{.2 1 0 .03}{#1}\pdfsetcolor{0 0 0 1}}
\def \dbvio#1{\pdfsetcolor{.2 1 0 .3}{#1}\pdfsetcolor{0 0 0 1}}
\def \mblue#1{\pdfsetcolor{1 1 0 .03}{#1}\pdfsetcolor{0 0 0 1}}
\def \dblue#1{\pdfsetcolor{1 1 0 .3}{#1}\pdfsetcolor{0 0 0 1}}

\def \ssqu{\rlap{\scalebox{.8}{$\blacksquare$}}\pdfsetcolor{0 0 0 .7}\scalebox{.8}{$\Box$}}
\def \sdia{\rlap{\scalebox{1.3}[.8]{$\blacklozenge$}}\pdfsetcolor{0 0 0 .7}\scalebox{1.4}[.8]{$\Diamond$}}
\def \stri{\rlap{\scalebox{1.2}[1.0]{$\blacktriangle$}}\pdfsetcolor{0 0 0 .7}\scalebox{1.2}[1.0]{$\vartriangle$}}
\def \sutr{\rlap{\scalebox{1.2}[1.0]{$\blacktriangledown$}}\pdfsetcolor{0 0 0 .7}\scalebox{1.2}[1.0]{$\triangledown$}}

\def \mcir{\rlap{\raisebox{-.1em}{\scalebox{1.8}{$\bullet$}}}\raisebox{-.1em}{\pdfsetcolor{0 0 0 .7}\scalebox{1.8}{$\circ$}}}
\def \msqu{\rlap{\scalebox{.95}{$\blacksquare$}}\pdfsetcolor{0 0 0 .7}\scalebox{.95}{$\Box$}}
\def \mdia{\rlap{\scalebox{1.5}[1.0]{$\blacklozenge$}}\pdfsetcolor{0 0 0 .7}\scalebox{1.6}[1.0]{$\Diamond$}}
\def \mtri{\rlap{\scalebox{1.4}[1.15]{$\blacktriangle$}}\pdfsetcolor{0 0 0 .7}\scalebox{1.4}[1.15]{$\vartriangle$}$\vphantom{|}$}
\def \mutr{\rlap{\scalebox{1.4}[1.15]{$\blacktriangledown$}}\pdfsetcolor{0 0 0 .7}\scalebox{1.4}[1.15]{$\triangledown$}$\vphantom{|}$}

\def \bsqu{\rlap{\scalebox{1.1}{$\blacksquare$}}\pdfsetcolor{0 0 0 .7}\scalebox{1.1}{$\Box$}}
\def \bdia{\rlap{\scalebox{1.8}[1.2]{$\blacklozenge$}}\pdfsetcolor{0 0 0 .7}\scalebox{1.8}[1.2]{$\Diamond$}}
\def \btri{\rlap{\scalebox{1.6}[1.5]{$\blacktriangle$}}\pdfsetcolor{0 0 0 .7}\scalebox{1.6}[1.5]{$\vartriangle$}$\vphantom{a^{\big(}}$}
\def \butr{\rlap{\scalebox{1.6}[1.5]{$\blacktriangledown$}}\pdfsetcolor{0 0 0 .7}\scalebox{1.6}[1.5]{$\triangledown$}$\vphantom{a^{\big(}}$}
\def \trip#1#2#3{\rlap{\raisebox{.15em}{\hspace{.2em}{#1}}}\rlap{\raisebox{-.2em}{#2}}\raisebox{-.2em}{\hspace{.4em}{#3}}}
\def \al{\alpha}
\def \be{\beta}
\def \ga{\gamma}
\def \de{\delta}
\def \ep{\epsilon}
\def \va{\varepsilon}

\def \th{\theta}

\def \la{\lambda}

\def \si{\sigma}
\def \ta{\tau}

\def \ph{\phi}

\def \om{\omega}
\def \Ga{\Gamma}

\def \La{\Lambda}

\def \Ph{\Phi}
\def \Ps{\Psi}

\def \pa{\partial}

\def \lb{\left[}
\def \rb{\right]}
\def \lp{\left(}
\def \rp{\right)}

\def \p#1{\phantom{#1}}
\def \vp#1{\vphantom{#1}}
\def \fr#1#2{{\textstyle \frac{#1}{#2}}}
\def \ha{\fr{1}{2}}
\def \nfr#1#2{\nicefrac{#1}{#2}}
\def \nha{\nfr{1}{2}}

\def \f#1{\underaccent{\_}{{#1}}}
\def \ff#1{\underaccent{=}{{#1}}}
\def \fff#1{\underaccent{\equiv}{{#1}}}
\def \nf#1{\underaccent{\sim}{{#1}}}
\def \ve#1{\accentset{\rightharpoonup}{{#1}}}
\def \ud#1{\underaccent{.}{{#1}}}
\def \udf#1{\underaccent{\_ .}{{#1}}}
\def \udff#1{\underaccent{= \!\p{.}\! \smash{\cdot}}{{#1}}}
\def \od#1{\accentset{\cdot}{{#1}}}

\title{An Exceptionally Simple Theory of Everything}

\author{A. Garrett Lisi\\
SLRI, 722 Tyner Way, Incline Village, NV 89451\\
E-mail: \email{alisi@hawaii.edu}}

\preprint{} 

\abstract{All fields of the standard model and gravity are unified as an E8 principal bundle connection. A non-compact real form of the E8 Lie algebra has G2 and F4 subalgebras which break down to strong su(3), electroweak su(2) x u(1), gravitational so(3,1), the frame-Higgs, and three generations of fermions related by triality. The interactions and dynamics of these 1-form and Grassmann valued parts of an E8 superconnection are described by the curvature and action over a four dimensional base manifold.}

\keywords{ToE}

\begin{document}
\setlength{\unitlength}{1mm} 


\section{Introduction}
We exist in a universe described by mathematics. But which math? Although it is interesting to consider that the universe may be the physical instantiation of all mathematics,\cite{Tegm} there is a classic principle for restricting the possibilities: The mathematics of the universe should be beautiful. A successful description of nature should be a concise, elegant, unified mathematical structure consistent with experience.

Hundreds of years of theoretical and experimental work have produced an extremely successful pair of mathematical theories describing our world. The standard model of particles and interactions described by quantum field theory is a paragon of predictive excellence. General relativity, a theory of gravity built from pure geometry, is exceedingly elegant and effective in its domain of applicability. Any attempt to describe nature at the foundational level must reproduce these successful theories, and the most sensible course towards unification is to extend them with as little new mathematical machinery as necessary. The further we drift from these experimentally verified foundations, the less likely our mathematics is to correspond with reality. In the absence of new experimental data, we should be very careful, accepting sophisticated mathematical constructions only when they provide a clear simplification. And we should pare and unite existing structures whenever possible.

The standard model and general relativity are the best mathematical descriptions we have of our universe. By considering these two theories and following our guiding principles, we will be led to a beautiful unification.

\subsection{A connection with everything}
\FIGURE[r]{
\epsfig{file=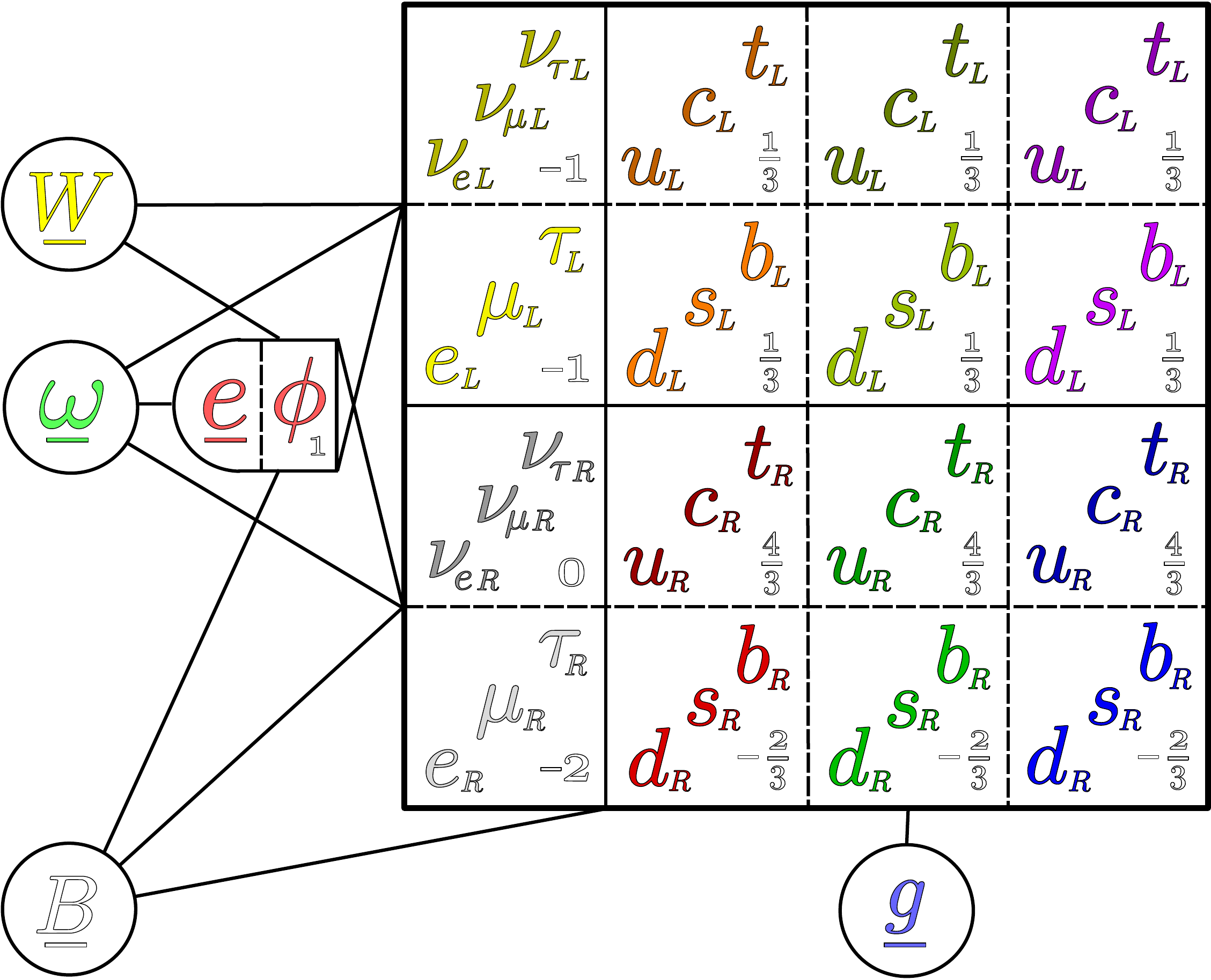, width=3.4in} $\!\!\!$
\caption{A periodic table of the standard model.$\protect\phantom{|_{\big|}}$ \label{ptsm}}}
\noindent
The building blocks of the standard model and gravity are fields over a four dimensional base manifold. The electroweak and strong gauge fields are described by Lie algebra valued connection 1-forms,   
$$
\f{W} \in \f{su}(2) \quad \f{B} \in \f{u}(1) \quad \f{g} \in \f{su}(3)
$$
while the gravitational fields are described by the spin connection,
$$
\f{\om} \in \f{so}(3,1) = \f{Cl}{}^2(3,1)
$$
a Clifford bivector valued 1-form, and the frame, $\f{e} \in \f{Cl}{}^1(3,1)$, a Clifford vector valued 1-form. The frame may be combined with a multiplet of Higgs scalar fields, $\ph$, to interact with the electroweak gauge fields and fermions to give them masses. The fermions are represented as Grassmann valued spinor fields, $\{ \ud{\nu}{}_e, \ud{e}, \ud{u}, \dots \}$, with the spin connection and gauge fields acting on them in fundamental representations. The electroweak $\f{W}$ acts on doublets of left chiral fermions, $\{ [ \ud{\nu}{}_{eL}, \ud{e}{}_L ], \dots \}$; the strong $\f{g}$ acts on triplets of red, green, and blue colored quarks, $\{ [ \ud{u}{}^r, \ud{u}{}^g, \ud{u}{}^b ], \dots \}$; and the electroweak $\f{B}$ acts on all with an interesting pattern of hypercharges. The left and right chiral parts of the gravitational spin connection, $\f{\om}$, act on the frame and on the left and right chiral fermions. This structure, depicted in Figure \ref{ptsm}, is repeated over three generations of fermions with different masses.

This diverse collection of fields in various algebras and representations is, inarguably, a mess. It is difficult at first to believe they can be unified as aspects of a unique mathematical structure --- but they can. The gauge fields are known to combine naturally as the connection of a grand unified theory with a larger Lie group, and we continue with unification in this spirit. The spin connection, frame, and Higgs may be viewed as Lie algebra elements and included as parts of a ``graviweak'' connection. Relying on the algebraic structure of the exceptional Lie groups, the fermions may also be recast as Lie algebra elements and included naturally as parts of a BRST extended connection.\cite{Lisi,Holt} The result of this program is a single principal bundle connection with everything,
\begin{equation}
\label{bigA}
\begin{array}{rcl}
\udf{A} &=& \ha \f{\om} + \fr{1}{4} \f{e} \ph + \f{B} + \f{W} + \f{g} + \vp{{}_\big(} \\
&& + \, \big( \ud{\nu}{}_e + \ud{e} + \ud{u} + \ud{d} \big)
+ \big( \ud{\nu}{}_\mu + \ud{\mu} + \ud{c} + \ud{s} \big)
+ \big( \ud{\nu}{}_\ta + \ud{\ta} + \ud{t} + \ud{b} \big)
\end{array}
\end{equation}
In this connection the bosonic fields, such as the strong $\f{g}=\f{dx^i} g_i^{\p{i}A} T_A$, are Lie algebra valued 1-forms, and the fermionic fields, such as $\ud{u} = \ud{u}{}^A T_A$, are Lie algebra valued Grassmann numbers.  (These Grassmann fields may be considered ghosts of former gauge fields, or accepted a priori as parts of this superconnection.)

The dynamics are described by the curvature,
\begin{equation}
\label{bigF}
\udff{F} = \f{d} \udf{A} + \ha [ \udf{A}, \udf{A} ]
\end{equation}
with interactions between particles given by their Lie bracket. For example, the interaction between two quarks and a gluon is specified by the Lie bracket between their generators, with a corresponding Feynman vertex,
$$
[ V_{g^{r\bar{g}}} , V_{u^g} ] = V_{u^r}
\qquad \Leftrightarrow \qquad
\parbox{20mm}{
\begin{picture}(20,20)
\put(0,0){
\begin{fmffile}{fmfv1} 
\begin{fmfgraph}(20,20)
\fmfleft{i1,i2}
\fmfright{o1}
\fmf{quark}{i1,v,o1}
\fmf{gluon}{i2,v}
\fmfdot{v}
\end{fmfgraph}
\end{fmffile}
}
\put(1,0){\mbox{$u^g$}}
\put(-1,17){\mbox{$g^{r\bar{g}}$}}
\put(20,6){\mbox{$u^r$}}
\end{picture}}
$$
It is a remarkable property of the exceptional Lie groups that some of their Lie brackets are equivalent to the action of a subgroup on vectors in fundamental representation spaces, just as they occur in the standard model.\cite{Adam} For example, the bracket between the gluons and a set of colored quarks in $\udf{A}$ can give the $su(3)$ action on the defining $3$,
$$
\Big[ \f{g}, \ud{u}{}^r + \ud{u}{}^g + \ud{u}{}^b \Big]
=
\f{g}
\lb
\begin{array}{c}
\ud{u}{}^r \\
\ud{u}{}^g \\
\ud{u}{}^b
\end{array}
\rb
$$
When all standard model particles and interactions are identified this way, the entire ensemble corresponds to a uniquely beautiful Lie group --- the largest simple exceptional group, $E8$.

\pagebreak

\section{The Standard Model Polytope}

The structure of a simple Lie algebra is described by its root system. An $N$ dimensional Lie algebra, considered as a vector space, contains an $R$ dimensional subspace, a \textbf{Cartan subalgebra}, spanned by a maximal set of $R$ inter-commuting generators, $T_a$,
$$
\big[ T_a, T_b \big] = T_a T_b - T_b T_a = 0 \qquad \forall \quad 1 \le a,b \le R
$$
($R$ is the \textbf{rank} of the Lie algebra) Every element of the Cartan subalgebra, $C=C^a T_a$, acts linearly on the rest of the Lie algebra via the Lie bracket (the adjoint action). The Lie algebra is spanned by the eigenvectors of this action, the \textbf{root vectors}, $V_\be$, with each corresponding to an eigenvalue,
$$
[ C , V_\be ] = \al_\be V_\be = \sum_a i C^a\al_{a\be} V_\be
$$
Each of the $(N\!-\!R)$ non-zero eigenvalues, $\al_\be$, (imaginary for real compact groups) is linearly dependent on the coefficients of $C$ and corresponds to a point, a \textbf{root}, $\al_{a\be}$, in the space dual to the Cartan subalgebra. The pattern of roots in $R$ dimensions uniquely characterizes the algebra and is independent of the choice of Cartan subalgebra and rotations of the constituent generators.

Since the root vectors, $V_\be$, and Cartan subalgebra generators, $T_a$, span the Lie algebra, they may be used as convenient generators --- the \textbf{Cartan-Weyl basis} of the Lie algebra,
$$
A = A^B T_B = A^a T_a + A^\be V_\be
$$
The Lie bracket between root vectors corresponds to vector addition between their roots, and to interactions between particles,
\begin{equation}
[ V_\be , V_\ga ] = V_\de \qquad \Leftrightarrow \qquad \al_{\be} + \al_{\ga} = \al_{\de} \qquad \Leftrightarrow \qquad
\parbox{20mm}{
\begin{picture}(20,20)
\put(0,0){
\begin{fmffile}{fmfv2} 
\begin{fmfgraph}(20,20)
\fmfleft{i1,i2}
\fmfright{o1}
\fmf{fermion}{i1,v,o1}
\fmf{photon}{i2,v}
\fmfdot{v}
\end{fmfgraph}
\end{fmffile}
}
\put(2,1){\mbox{$\ga$}}
\put(2,17){\mbox{$\be$}}
\put(22,6){\mbox{$\de$}}
\end{picture}}
\vp{\Big(_{\Big(}^{\Big(}}
\label{interaction}
\end{equation}
Elements of the Lie algebra and Cartan subalgebra can also act on vectors in the various representation spaces of the group. In these cases the eigenvectors of the Cartan subalgebra (called \textbf{weight vectors}) have eigenvalues corresponding to the generalized roots (called \textbf{weights}) describing the representation. From this more general point of view, the roots are the weights of the Lie algebra elements in the adjoint representation space. 

Each weight vector, $V_\be$, corresponds to a type of elementary particle. The $R$ coordinates of each weight are the quantum numbers of the relevant particle with respect to the chosen Cartan subalgebra generators.

\subsection{Strong $G2$}

\begin{floatingtable}
{\centerline{\parbox{.45\textwidth}{\centerline{\begin{tabular}
{@{\vrule width1.0pt}c@{\vrule width0.2pt}c@{\vrule width1.0pt}c@{\vrule width1.0pt}c@{\vrule width0.2pt}c@{\vrule width1.0pt}}\noalign{\hrule height 1.0pt}
\multicolumn{2}{@{\vrule width1.0pt}c@{\vrule width1.0pt}}{$G2$} & $V_\be$ & $\;\, g^3 \,\;$ & $\;\, g^8 \,\;$\\
\noalign{\hrule height 1pt}
\,\lblue{\mcir}\, & $\, g^{r\bar{g}} \,$ & $(T_2 - i T_1)$ & $1$ & $0$ \\
\lblue{\mcir} & $g^{\bar{r}g}$ & $\, (-T_2 - i T_1) \,$ & $-1$ & $0$ \\
\lblue{\mcir} & $g^{r\bar{b}}$ & $\, (T_5 - i T_4) \,$ & $\nha$ & $\nfr{\sqrt{3}}{2}$ \\
\lblue{\mcir} & $\, g^{\bar{r}b} \,$ & $\, (-T_5 - i T_4) \,$ & $\nfr{-1}{2}$ & $\nfr{-\sqrt{3}}{2}$ \\
\lblue{\mcir} & $g^{\bar{g}b}$ & $(-T_7 - i T_6)$ & $\nha$ & $\nfr{-\sqrt{3}}{2}$ \\
\lblue{\mcir} & $g^{g\bar{b}}$ & $(T_7 - i T_6)$ & $\nfr{-1}{2}$ & $\nfr{\sqrt{3}}{2}$ \\
\noalign{\hrule height 0.2pt}
\mred{\mtri} & $q^r$ & $[1,0,0]$ & $\nha$ & $\nfr{1}{2\sqrt{3}}$ \\
\mred{\mutr} & ${\bar q}{}^r$ & $[1,0,0]$ & $\, \nfr{-1}{2} \,$ & $\, \nfr{-1}{2\sqrt{3}} \,$ \\
\mgree{\mtri} & $q^g$ & $[0,1,0]$ & $\nfr{-1}{2}$ & $\nfr{1}{2\sqrt{3}}$ \\
\mgree{\mutr} & ${\bar q}{}^g$ & $[0,1,0]$ & $\nha$ & $\nfr{-1}{2\sqrt{3}}$ \\
\mblue{\mtri} & $q^b$ & $[0,0,1]$ & $0$ & $\nfr{-1}{\sqrt{3}}$ \\
\mblue{\mutr} & ${\bar q}{}^b$ & $[0,0,1]$ & $0$ & $\nfr{1}{\sqrt{3}}$ \\
\noalign{\hrule height 1.0pt}
\end{tabular}
		}}~~~~
		\parbox{.45\textwidth}{\centerline{
		\epsfig{file=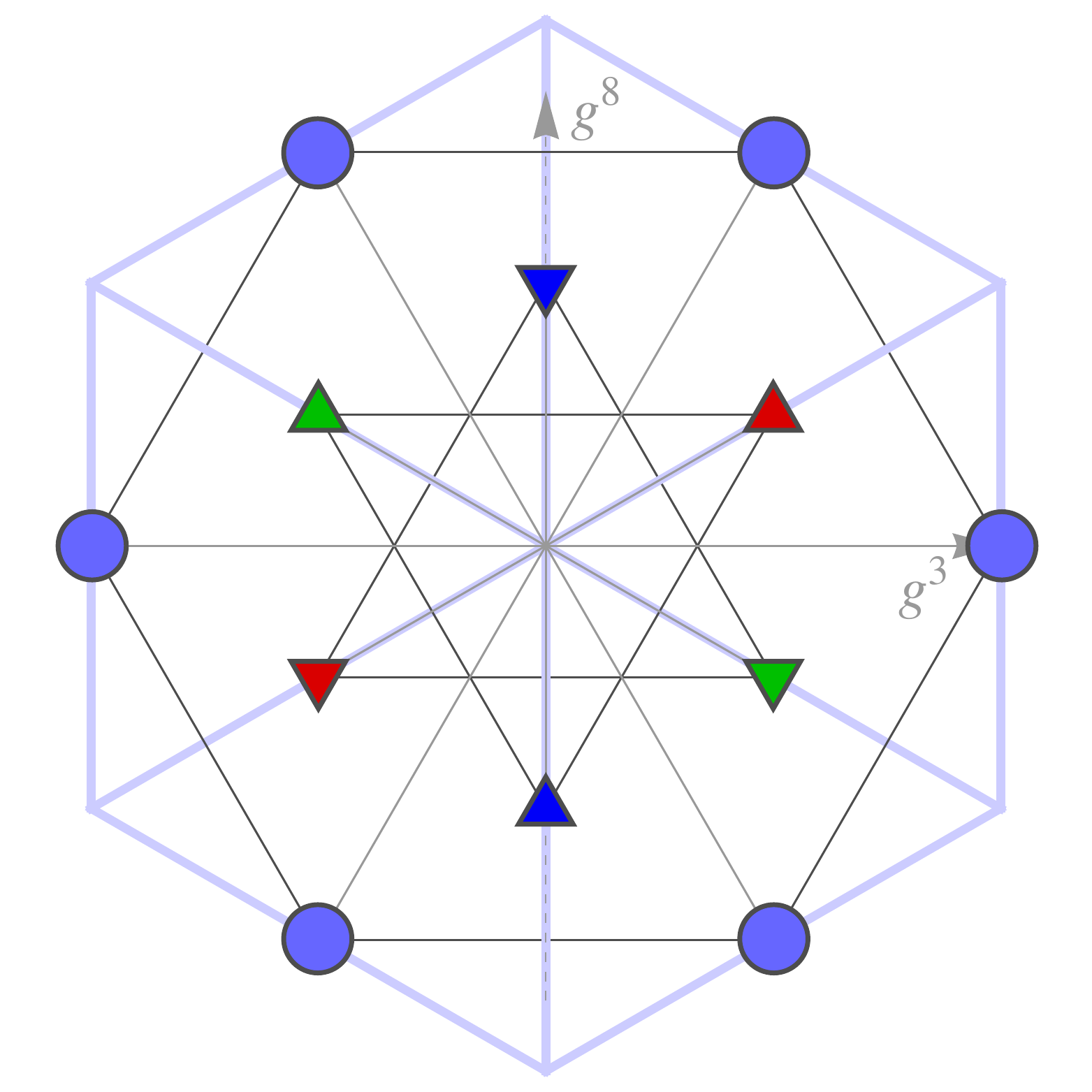, width=3in}
			}}}
\caption{The $su(3)$ weight vectors and weight coordinates of the gluon, quark, and anti-quark weights form the G2 root system.\label{g2t}}}
\end{floatingtable}
\noindent
The gluons, $\f{g} \in \f{su}(3)$, in the special unitary group of degree three may be represented using the eight Gell-Mann matrices as generators,
\begin{equation}
\label{gluons}
\begin{array}{rcl}
g &=& g^A T_A =  g^A \fr{i}{2} \la_A
= C + g^\be V_\be = \vp{{}_{\big(}} \\
&=&
\fr{i}{2}
\lb
\begin{array}{ccc}
 g^3 \!+\! \fr{1}{\sqrt{3}} g^8 \! & g^1\!-\!ig^2 & g^4\!-\!ig^5 \\
g^1\!+\!ig^2 & \! -g^3 \!+\! \fr{1}{\sqrt{3}} g^8 & g^6\!-\!ig^7 \\
g^4\!+\!ig^5 & g^6\!+\!ig^7 & \fr{-2}{\sqrt{3}} g^8
\end{array}
\rb
=
\lb
\begin{array}{ccc}
 \fr{i}{2} g^3 \!+\! \fr{i}{2\sqrt{3}} g^8 \! & g^{r\bar{g}} & g^{r\bar{b}} \\
g^{\bar{r}g} & \! \fr{-i}{2} g^3 \!+\! \fr{i}{2\sqrt{3}} g^8 & g^{g\bar{b}} \\
g^{\bar{r}b} & g^{\bar{g}b} & \fr{-i}{\sqrt{3}} g^8
\end{array}
\rb
\end{array}
\end{equation}
The Cartan subalgebra, $C = g^3 T_3 + g^8 T_8$, is identified with the diagonal. This gives root vectors --- particle types --- corresponding to the six non-zero roots, such as
$$
\begin{array}{c}
\big[ C , V_{g^{g\bar{b}}} \big] = i \lp g^3 \lp \nfr{-1}{2} \rp  + g^8 ( \nfr{\sqrt{3}}{2} ) \rp V_{g^{g\bar{b}}} \vp{|_\Big(} \\
V_{g^{g\bar{b}}} = \lp T_7-iT_6 \rp =
\lb \matrix{
0 & 0 & 0 \cr
0 & 0 & 1 \cr
0 & 0 & 0 \cr
} \rb
\qquad
g^{g\bar{b}} = g^{g\bar{b}} V_{g^{g\bar{b}}} = \fr{i}{2} ( g^6 - i g^7 ) V_{g^{g\bar{b}}} =
\lb \matrix{
0 & 0 & 0 \cr
0 & 0 & g^{g\bar{b}} \cr
0 & 0 & 0 \cr
} \rb
\end{array}
$$
for the green anti-blue gluon. (By an abuse of notation, the coefficient, such as $g^{g\bar{b}}$, has the same label as the particle eigenvector containing the coefficient, and as the root --- the usage is clear from context.)

Since the Cartan subalgebra matrix in the standard representation acting on $3$, and its dual acting on $\bar{3}$, are diagonal, the weight vectors, $V_\be$ and $\bar{V}{}_\be$, satisfying
$$
C V_\be = \sum_a i g^a\al_{a\be} V_\be
\qquad {\rm and} \qquad
\bar{C} \bar{V}{}_\be = -C^T \bar{V}{}_\be = \sum_a i g^a \al_{a\be} \bar{V}{}_\be
$$
are the canonical unit vectors of the $3$ and $\bar{3}$. The weights for these --- the $su(3)$ quantum numbers of the quarks and anti-quarks --- can be read off the diagonals of $C$ and $\bar{C} = -C^T = -C$.

The set of weights for $su(3)$, the defining $3$, and its dual $\bar{3}$, are shown in Table \ref{g2t}. These weights are precisely the $12$ roots of the rank two simple exceptional Lie group, $G2$. The weight vectors and weights of the $3$ and $\bar{3}$ are identified as root vectors and roots of $G2$. The $G2$ Lie algebra breaks up as
$$
g2 = su(3) + 3 + \bar{3}
$$
allowing a connection to be separated into the $\f{su}(3)$ gluons, $\f{g}$, and the $3$ and $\bar{3}$ quarks and anti-quarks, $\ud{q}$ and $\ud{\bar{q}}$, related by Lie algebra duality. All interactions (\ref{interaction}) between gluons and quarks correspond to vector addition of the roots of $G2$, such as
$$ \!\!\!\!\!\!\!
[ V_{g^{r\bar{g}}} , V_{q^g} ] = V_{q^r}
\;\;\, \Leftrightarrow \;\;\,
g^{r\bar{g}} + q^g = q^r
\;\;\, \Leftrightarrow \;\;\,
(1,0) + (\fr{-1}{2}, \fr{1}{2\sqrt{3}}) = (\ha, \fr{1}{2\sqrt{3}})
\;\;\, \Leftrightarrow \;\;
\parbox{20mm}{
\begin{picture}(20,20)
\put(0,0){
\begin{fmffile}{fmfv3} 
\begin{fmfgraph}(20,20)
\fmfleft{i1,i2}
\fmfright{o1}
\fmf{quark}{i1,v,o1}
\fmf{gluon}{i2,v}
\fmfdot{v}
\end{fmfgraph}
\end{fmffile}
}
\put(2,1){\mbox{$q^g$}}
\put(-1,17){\mbox{$g^{r\bar{g}}$}}
\put(20,6){\mbox{$q^r$}}
\end{picture}}
\vp{{\Big|}_{\Big|_\big|}^{\Big|^\big|}}
$$
We are including these quarks in a simple exceptional Lie algebra, $g2$, and not merely acting on them with $su(3)$ in some representation. The necessity of specifying a representation for the quarks has been removed --- a significant simplification of mathematical structure. And we will see that this simplification does not occur only for the quarks in $g2$, but for all fermions of the standard model. 

Just as we represented the gluons in the $(3 \times 3)$ matrix representation (\ref{gluons}) of $su(3)$, we may choose to represent the gluons and quarks using the smallest irreducible, $(7 \times 7)$, matrix representation of $g2$,\cite{Cacc}
\begin{equation}
g + q + \bar{q}
=
{\textstyle
\lb \begin{array}{ccccccc}
0 & \fr{-1}{\sqrt{2}} \bar{q}^b & \fr{-1}{\sqrt{2}} q^b & \fr{-1}{\sqrt{2}} q^r & \fr{-1}{\sqrt{2}} \bar{q}^r & \fr{-1}{\sqrt{2}} q^g & \fr{-1}{\sqrt{2}} \bar{q}^g \\
\fr{1}{\sqrt{2}} q^b & \fr{i}{\sqrt{3}} g^8 & 0 & \fr{1}{\sqrt{2}} \bar{q}^g & g^{\bar{r}b} & \fr{-1}{\sqrt{2}} \bar{q}^r & -g^{\bar{g}b} \\ 
\fr{1}{\sqrt{2}} \bar{q}^b & 0 & \fr{-i}{\sqrt{3}} g^8 & g^{r\bar{b}} & \fr{1}{\sqrt{2}} q^g & -g^{g\bar{b}} & \fr{-1}{\sqrt{2}} q^r \\ 
\fr{1}{\sqrt{2}} \bar{q}^r & \fr{-1}{\sqrt{2}} q^g & -g^{\bar{r}b} & \fr{i}{2} g^3 \!+\! \fr{i}{2\sqrt{3}} g^8 \! & 0 & g^{\bar{r}g} & \fr{1}{\sqrt{2}} q^b \\ 
\fr{1}{\sqrt{2}} q^r & -g^{r\bar{b}} & \fr{-1}{\sqrt{2}} \bar{q}^g & 0 & \! \fr{-i}{2} g^3 \!-\! \fr{i}{2\sqrt{3}} g^8 \! & \fr{1}{\sqrt{2}} \bar{q}^b & g^{r\bar{g}} \\ 
\fr{1}{\sqrt{2}} \bar{q}^g & \fr{1}{\sqrt{2}} q^r & g^{\bar{g}b} & -g^{r\bar{g}} & \fr{-1}{\sqrt{2}} q^b & \! \fr{-i}{2} g^3 \!+\! \fr{i}{2\sqrt{3}} g^8 \! & 0 \\ 
\fr{1}{\sqrt{2}} q^g &  g^{g\bar{b}} & \fr{1}{\sqrt{2}} \bar{q}^r & \fr{-1}{\sqrt{2}} \bar{q}^b & -g^{\bar{r}g} & 0 & \! \fr{i}{2} g^3 \!-\! \fr{i}{2\sqrt{3}} g^8  
\end{array} \rb
}
\label{7rep}
\end{equation}
Squaring this matrix gives all interactions between gluons and quarks, equivalent to $su(3)$ acting on quarks and anti-quarks in the fundamental representation spaces.

\begin{floatingtable}{\centerline{\parbox{.45\textwidth}{\centerline{
\begin{tabular}
{@{\vrule width1.0pt}c@{\vrule width0.2pt}c@{\vrule width1.0pt}c@{\vrule width0.2pt}c@{\vrule width0.2pt}c@{\vrule width1.0pt}c@{\vrule width0.2pt}c@{\vrule width0.2pt}c@{\vrule width1.0pt}}
\noalign{\hrule height 1.0pt}
\multicolumn{2}{@{\vrule width1.0pt}c@{\vrule width1.0pt}}{$\, G2+U(1) \,$} & $\;\; x \;\;$ & $\;\; y \;\;$ & $\;\; z \;\;$ & $\, \fr{\sqrt{2}}{\sqrt{3}} B_2 \,$ & $\;\, g^3 \,\;$ & $\;\, g^8 \,\;$ \\
\noalign{\hrule height 1.0pt}
\lblue{\mcir} & $g^{r\bar{g}}$ & $-1$ & $1$& $0$ & $0$ & $1$ & $0$ \\
\lblue{\mcir} & $g^{\bar{r}g}$ & $1$ & $-1$& $0$ & $0$ & $-1$ & $0$ \\
\lblue{\mcir} & $g^{r\bar{b}}$ & $-1$ & $0$& $1$ & $0$ & $\nfr{1}{2}$ & $\nfr{\sqrt{3}}{2}$ \\
\, \lblue{\mcir} \, & $\; g^{\bar{r}b} \;$ & $1$ & $0$& $-1$ & $0$ & $\nfr{-1}{2}$ & $\nfr{-\sqrt{3}}{2}$ \\
\lblue{\mcir} & $g^{\bar{g}b}$ & $0$ & $1$& $-1$ & $0$ & $\nfr{1}{2}$ & $\nfr{-\sqrt{3}}{2}$ \\
\lblue{\mcir} & $g^{g\bar{b}}$ & $0$ & $-1$& $1$ & $0$ & $\nfr{-1}{2}$ & $\nfr{\sqrt{3}}{2}$ \\
\noalign{\hrule height 1.0pt}
\mred{\btri} & $q^r_{I}$ & $\nfr{-1}{2}$ & $\nfr{1}{2}$ & $\nfr{1}{2}$ & $\nfr{-1}{6}$ & $\nfr{1}{2}$ & $\nfr{1}{2\sqrt{3}}$ \\
\mgree{\btri} & $q^g_{I}$ & $\nfr{1}{2}$ & $\nfr{-1}{2}$ & $\nfr{1}{2}$ & $\, \nfr{-1}{6} \,$ & $\, \nfr{-1}{2} \,$ & $\nfr{1}{2\sqrt{3}}$ \\
\mblue{\btri} & $q^b_{I}$ & $\nfr{1}{2}$ & $\nfr{1}{2}$ & $\nfr{-1}{2}$ & $\nfr{-1}{6}$ & $0$ & $\nfr{-1}{\sqrt{3}}$ \\
\noalign{\hrule height 0.2pt}
\mred{\butr} & ${\bar q}{}^r_{I}$ & $\nfr{1}{2}$ & $\nfr{-1}{2}$ & $\nfr{-1}{2}$ & $\nfr{1}{6}$ & $\nfr{-1}{2}$ & $\nfr{-1}{2\sqrt{3}}$ \\
\mgree{\butr} & ${\bar q}{}^g_{I}$ & $\nfr{-1}{2}$ & $\nfr{1}{2}$ & $\nfr{-1}{2}$ & $\nfr{1}{6}$ & $\nfr{1}{2}$ & $\, \nfr{-1}{2\sqrt{3}} \,$ \\
\mblue{\butr} & ${\bar q}{}^b_{I}$ & $\nfr{-1}{2}$ & $\nfr{-1}{2}$ & $\nfr{1}{2}$ & $\nfr{1}{6}$ & $0$ & $\nfr{1}{\sqrt{3}}$ \\
\noalign{\hrule height 1.0pt}
\mgray{\btri} & $l$ & $\, \nfr{-1}{2} \,$ & $\, \nfr{-1}{2} \,$ & $\, \nfr{-1}{2} \,$ & $\nfr{1}{2}$ & $0$ & $0$ \\
\noalign{\hrule height 0.2pt}
\mgray{\butr} & ${\bar l}$ & $\nfr{1}{2}$ & $\nfr{1}{2}$ & $\nfr{1}{2}$ & $\nfr{-1}{2}$ & $0$ & $0$ \\
\noalign{\hrule height 1.0pt}
\end{tabular}}}~~~~
\parbox{.45\textwidth}{\centerline{
\begin{tabular}
{@{\vrule width1.0pt}c@{\vrule width0.2pt}c@{\vrule width1.0pt}c@{\vrule width0.2pt}c@{\vrule width0.2pt}c@{\vrule width1.0pt}c@{\vrule width0.2pt}c@{\vrule width0.2pt}c@{\vrule width1.0pt}}
\noalign{\hrule height 1.0pt}
\multicolumn{2}{@{\vrule width1.0pt}c@{\vrule width1.0pt}}{$\, G2+U(1) \,$} & $\;\; x \;\;$ & $\;\; y \;\;$ & $\;\; z \;\;$ & $\, \fr{\sqrt{2}}{\sqrt{3}} B_2 \,$ & $\;\, g^3 \,\;$ & $\;\, g^8 \,\;$ \\
\noalign{\hrule height 1pt}
\mred{\stri} & $q^r_{II}$ & $-1$ & $0$& $0$ & $\nfr{1}{3}$ & $\nfr{1}{2}$ & $\nfr{1}{2\sqrt{3}}$ \\
\mgree{\stri} & $q^g_{II}$ & $0$ & $-1$& $0$ & $\, \nfr{1}{3} \,$ & $\, \nfr{-1}{2} \,$ & $\nfr{1}{2\sqrt{3}}$ \\
\, \mblue{\stri} \, & $\, q^b_{II} \,$ & $0$ & $0$& $-1$ & $\nfr{1}{3}$ & $0$ & $\nfr{-1}{\sqrt{3}}$ \\
\noalign{\hrule height 0.2pt}
\mred{\sutr} & ${\bar q}{}^r_{II}$ & $1$ & $0$& $0$ & $\nfr{-1}{3}$ & $\nfr{-1}{2}$ & $\nfr{-1}{2\sqrt{3}}$ \\
\mgree{\sutr} & ${\bar q}{}^g_{II}$ & $0$ & $1$& $0$ & $\nfr{-1}{3}$ & $\nfr{1}{2}$ & $\, \nfr{-1}{2\sqrt{3}} \,$ \\
\mblue{\sutr} & ${\bar q}{}^b_{II}$ & $0$ & $0$& $1$ & $\nfr{-1}{3}$ & $0$ & $\nfr{1}{\sqrt{3}}$ \\
\noalign{\hrule height 1.0pt}
\mred{\msqu} & $q^r_{III}$ & $0$ & $1$& $1$ & $\nfr{-2}{3}$ & $\nfr{1}{2}$ & $\nfr{1}{2\sqrt{3}}$ \\
\mgree{\msqu} & $q^g_{III}$ & $1$ & $0$& $1$ & $\, \nfr{-2}{3} \,$ & $\, \nfr{-1}{2} \,$ & $\nfr{1}{2\sqrt{3}}$ \\
\mblue{\msqu} & $q^b_{III}$ & $1$ & $1$& $0$ & $\nfr{-2}{3}$ & $0$ & $\nfr{-1}{\sqrt{3}}$ \\
\noalign{\hrule height 0.2pt}
\mred{\mdia} & ${\bar q}{}^r_{III}$ & $0$ & $-1$& $-1$ & $\nfr{2}{3}$ & $\nfr{-1}{2}$ & $\nfr{-1}{2\sqrt{3}}$ \\
\mgree{\mdia} & ${\bar q}{}^g_{III}$ & $-1$ & $0$& $-1$ & $\nfr{2}{3}$ & $\nfr{1}{2}$ & $\, \nfr{-1}{2\sqrt{3}} \,$ \\
\mblue{\mdia} & ${\bar q}{}^b_{III}$ & $-1$ & $-1$& $0$ & $\nfr{2}{3}$ & $0$ & $\nfr{1}{\sqrt{3}}$ \\
\noalign{\hrule height 1.0pt}
\end{tabular}}}}
\caption{Weights of gluons, three series of quarks and anti-quarks, and leptons, in three dimensions, projecting down to the $G2$ root system in the last two coordinates.\label{g23dt}}}
\end{floatingtable}

The $G2$ root system may also be described in three dimensions as the $12$ midpoints of the edges of a cube --- the vertices of a cuboctahedron. These roots are labeled $g$ and $q_{III}$  in Table \ref{g23dt}, with their $(x,y,z)$ coordinates shown. These points may be rotated and scaled,
\begin{equation}
\label{su3rot}
\lb
\begin{array}{c}
B_2 \\
g^3 \\
g^8
\end{array}
\rb
=
\frac{1}{\sqrt{2}}
\lb
\begin{array}{ccc}
\fr{-1}{\sqrt{3}} & \fr{-1}{\sqrt{3}} & \fr{-1}{\sqrt{3}}\\
\fr{-1}{\sqrt{2}} & \fr{1}{\sqrt{2}} & 0 \\
\fr{-1}{\sqrt{6}} & \fr{-1}{\sqrt{6}} & \fr{\sqrt{2}}{\sqrt{3}}
\end{array}
\rb
\lb
\begin{array}{c}
x \\
y \\
z
\end{array}
\rb
\end{equation}
so that dropping the first, $B_2$, coordinate gives the projection to the roots in two dimensions.

In general, we can find subalgebras by starting with the root system of a Lie algebra, rotating it until multiple roots match up on parallel lines, and collapsing the root system along these lines to an embedded space of lower dimension --- a projection. Since the cuboctahedron is the root system of $so(6)$, we have obtained $g2$ by projecting along a $u(1)$ in the Cartan subalgebra of $so(6)$,
\begin{equation}
so(6) = su(4) = u(1) + su(3) + 3 + \bar{3} \to u(1) + g2
\label{su4}
\end{equation}
This particular rotation and projection (\ref{su3rot}) generalizes to give the $su(n)$ subalgebra of any $so(2n)$. We can also obtain $g2$ as a projected subalgebra of $so(7)$ --- the root system is the $so(6)$ root system plus $6$ shorter roots, labeled $q_{II}$, at the centers of the faces of the cube in the figure of Table \ref{g2t}. The eight weights at the corners of a half-cube, labeled $q_I$ and $l$, also project down to the roots of $G2$ and the origin, giving leptons and anti-leptons in addition to quarks,
\begin{equation}
su(3) + 3 + \bar{3} + 1 + \bar{1}
\label{su31}
\end{equation}
These three series of weights in three dimensions, and their rotations into $su(3)$ coordinates, are shown in Table \ref{g23dt}. The action of $su(3)$ on quarks and leptons corresponds to its action on these sets of weights, while the $u(1)_{B-L}$ quantum number, $B_2$, is the baryon minus lepton number, related to their hypercharge. The $su(3)$ action does not move fermions between the nine $B_2$ grades in the table --- each remains in its series, $I$, $II$, or $III$. Since this $su(3)$ and $u(1)_{B-L}$ are commuting subalgebras, our grand unification of gauge fields follows the same path as the Pati-Salam $SU(2)_L \times SU(2)_R \times SU(4)$ GUT.\cite{Pati}

\subsection{Graviweak $F4$}

The interactions between other gauge fields are more involved and separate from the strong gluons. Most importantly, the weak $\f{W}$ acts only on left-chiral fermions, as determined by their gravitational $so(3,1)$ quantum numbers. Also, the Higgs, $\ph$, needs to be combined with the gravitational frame, $\f{e}$, to make a 1-form interacting correctly with the electroweak gauge fields and the fermions. These interactions imply that the spin connection, which acts on the frame, and the electroweak gauge fields, which act on the Higgs, must be combined in a graviweak gauge group. The best candidate for this unification is $so(7,1)$, which breaks up as
\begin{equation}
so(7,1) = so(3,1) + so(4) + (4 \times 4) = so(3,1) + \big( su(2)_L + su(2)_R \big) + \big( 4 \times (2+\bar{2}) \big)
\label{so71algebrabreakdown}
\end{equation}
and has the desired balance of gravity and left-right symmetric electroweak gauge fields acting on the frame-Higgs.

\subsubsection{Gravitational $D2$}

For its action on spinors, gravity is best described using the spacetime Clifford algebra, $Cl(3,1)$ --- a Lie algebra with a symmetric product. The four orthonormal Clifford vector generators,
$$
\ga_1 = \si_2 \otimes \si_1 \qquad
\ga_2 = \si_2 \otimes \si_2 \qquad
\ga_3 = \si_2 \otimes \si_3 \qquad
\ga_4 = i \si_1 \otimes 1
$$
are written here as $(4\times4)$ Dirac matrices in a chiral representation, built using the Kronecker product of Pauli matrices,
$$
\si_1 =
\lb \matrix{
0 & 1 \cr
1 & 0
} \rb
\qquad
\si_2 =
\lb \matrix{
0 & \!\!\!-i \cr
i & 0
} \rb
\qquad
\si_3 = 
\lb \matrix{
1 & 0 \cr
0 & \!\!\!-1
} \rb
$$
These may be used to write the gravitational frame as
$$
\f{e} = \f{dx^i} (e_i)^\mu \ga_\mu
=
i
\lb \matrix{
0 & \f{e}{}^4 \!-\! \f{e}{}^\va \si_\va \cr
\f{e}{}^4 \!+\! \f{e}{}^\va \si_\va & 0
} \rb
=
\lb \matrix{
0 & \f{e}{}_R \cr
\f{e}{}_L & 0
} \rb
=
\lb \matrix{
0 & 0 & \f{e}{}_T^\vee & \f{e}{}_S^\wedge \cr
0 & 0 & \f{e}{}_S^\vee & \f{e}{}_T^\wedge \cr
\f{e}{}_T^\wedge & -\f{e}{}_S^\wedge & 0 & 0 \cr
-\f{e}{}_S^\vee & \f{e}{}_T^\vee & 0 & 0
} \rb
$$
with left and right chiral parts, $\f{e}{}_{L/R}= i(\f{e}{}^4 \!\pm\! \f{e}{}^\va \si_\va)$, and the coefficients,
$$
\begin{array}{rclcrcl}
\f{e}{}_T^\wedge &=& i \f{e}{}^4 \!+\! i \f{e}{}^3 &\qquad& \f{e}{}_S^\wedge &=& -i \f{e}{}^1 \!-\! \f{e}{}^2 \\
\f{e}{}_T^\vee &=& i \f{e}{}^4 \!-\! i \f{e}{}^3 &\qquad& \f{e}{}_S^\vee &=& -i \f{e}{}^1 \!+\! \f{e}{}^2 = -\f{e}{}_S^{\wedge*}
\end{array}
$$
The $d2 = so(3,1) = Cl^2(3,1)$ valued gravitational spin connection is written using the six Clifford bivector generators, $\ga_{\mu \nu} = \ha [ \ga_\mu, \ga_\nu ]$, as
\begin{equation}
\label{so31}
\begin{array}{rcl}
\f{\om} &=& \ha \f{\om}{}^{\mu \nu} \ga_{\mu \nu}
=
\lb \begin{array}{cc}
( \fr{1}{2} \f{\om}^{\va \pi} \ep_{\va \pi}^{\p{\va \pi} \ta} \!-\! i \f{\om}^{\ta 4} ) i \si_\ta & 0 \\
0 & ( \fr{1}{2} \f{\om}^{\va \pi} \ep_{\va \pi}^{\p{\va \pi} \ta} \!+\! i \f{\om}^{\ta 4} ) i \si_\ta
\end{array} \rb
=
\vp{\Big|_{\Big|_\Big|}}
\\
&=&
\lb \begin{array}{cc}
 ( \f{\om}{}_S^\ta \!-\! i \f{\om}{}_T^\ta ) i \si_\ta & 0 \\
0 & ( \f{\om}{}_S^\ta \!+\! i \f{\om}{}_T^\ta ) i \si_\ta
\end{array} \rb
=
\lb \begin{array}{cc}
\f{\om}{}_L & 0 \\
0 & \f{\om}{}_R
\end{array} \rb
\end{array}
\end{equation}

\TABLE[r]{
\begin{tabular}{@{\vrule width1.0pt}c@{\vrule width0.2pt}c@{\vrule width1.0pt}c@{\vrule width0.2pt}c@{\vrule width1.0pt}c@{\vrule width0.2pt}c@{\vrule width1.0pt}}\noalign{\hrule height 1.0pt}
\multicolumn{2}{@{\vrule width1.0pt}c@{\vrule width1.0pt}}{$D2_G$} & $\, \fr{1}{2i} \om_T^3 \,$ & $\, \ha \om_S^3 \,$ & $\, \ha \om_L^3 \,$ & $\, \ha \om_R^3 \,$ \\
\noalign{\hrule height 1pt}
\lgree{\mcir} & $\, \om_L^\wedge \,$ & $1$ & $1$ & $1$ & $0$ \\
\lgree{\mcir} & $\om_L^\vee$ & $\, -1 \,$ & $-1$ & $\, -1 \,$ & $0$  \\
\lgree{\mcir} & $\om_R^\wedge$ & $-1$ & $1$ & $0$ & $1$ \\
\lgree{\mcir} & $\om_R^\vee$ & $1$ & $-1$ & $0$ & $-1$ \\
\noalign{\hrule height 1.0pt}
\lred{\msqu} & $e_S^\wedge$ & $0$ & $1$ & $\nha$ & $\nha$ \\
\lred{\msqu} & $e_S^\vee$ & $0$ & $-1$ & $\nfr{-1}{2}$ & $\nfr{-1}{2}$ \\
\, \lred{\msqu} \, & $e_T^\wedge$ & $1$ & $0$ & $\nfr{-1}{2}$ & $\nha$ \\
\lred{\msqu} & $e_T^\vee$ & $-1$ & $0$ & $\nha$ & $\nfr{-1}{2}$ \\
\noalign{\hrule height 1.0pt}
\myell{\mtri} & $f_L^\wedge$ & $\nha$ & $\nha$ & $\nha$ & $0$ \\
\myell{\mtri} & $f_L^\vee$ & $\nfr{-1}{2}$ & $\nfr{-1}{2}$ & $\, \nfr{-1}{2} \,$ & $0$ \\
\mgray{\mtri} & $f_R^\wedge$ & $\nfr{-1}{2}$ & $\nha$ & $0$ & $\nha$ \\
\mgray{\mtri} & $f_R^\vee$ & $\nha$ & $\, \nfr{-1}{2} \,$ & $0$ & $\, \nfr{-1}{2} \,$ \\
\noalign{\hrule height 1.0pt}
\end{tabular} $\!\!\!\!\!$
\caption{Gravitational $D2$ weights for the spin connection, frame, and fermions, in two coordinate systems. \label{gd2t}}}
\noindent
with six real coefficients redefined into the spatial rotation and temporal boost parts,
$$
\f{\om}{}_S^\ta = \fr{1}{2} \f{\om}^{\va \pi} \ep_{\va \pi}^{\p{\va \pi} \ta}
\qquad
\f{\om}{}_T^\ta = \f{\om}^{\ta 4}
$$
These relate to the left and right-chiral (selfdual and anti-selfdual) parts of the spin connection,
$$
\f{\om}{}_{L/R} = ( \f{\om}{}_{L/R}^\ta ) i \si_\ta =  \f{\om}{}_S \mp i \f{\om}{}_T
$$
which are $sl(2,\mathbb{C})$ valued but not independent, $\f{\om}{}_R^\ta=\f{\om}{}_L^{\ta*}$.

The Cartan subalgebra of gravity, in several different coordinates, is
\begin{eqnarray*}
C &=& \om^{12} \ga_{12} + \om^{34} \ga_{34}
= \om_S^3 \ga_{12} + \om_T^3 \ga_{34} = 
\vp{{}_{\big|}}
\\
&=&
\lb \begin{array}{cc}
 ( \f{\om}{}_S^3 \!-\! i \f{\om}{}_T^3 ) i \si_3 & 0 \\
0 & ( \f{\om}{}_S^3 \!+\! i \f{\om}{}_T^3 ) i \si_3
\end{array} \rb= 
\lb \begin{array}{cc}
\om_L^3 i \si_3 & 0 \\
0 & \om_R^3 i \si_3
\end{array} \rb
\end{eqnarray*}
Taking the Lie bracket with $C$ gives root vectors and roots for the spin connection, such as
$$
\qquad \quad
\lb C , \fr{1}{4}( - \ga_{13} + \ga_{14} - i \ga_{23} + i \ga_{24} ) \rb = 
i \lp \om^3_S (2) + \fr{1}{i} \om^3_T (2) \rp \fr{1}{4}( - \ga_{13} + \ga_{14} - i \ga_{23} + i \ga_{24} )
$$
for $\om_L^\wedge$, and weight vectors and weights for the frame, such as
$$
\lb C , \fr{i}{2} (\ga_3 - \ga_4 ) \rb
= i \lp \fr{1}{i} \om^3_T (2) \rp \fr{i}{2} (\ga_3 - \ga_4 )
$$
for $e_T^\wedge$. The fermions, such as the left-chiral spin-up up quark, $\ud{u}{}_L^\wedge$, are in the $4$ of the spinor representation space (\ref{so31}) with weight vectors, such as $[1,0,0,0]$, equal to the canonical unit vectors, and weights read off the diagonal of $C$. The collection of fields and their weights are shown in Table \ref{gd2t}. The two coordinate systems in the table are related by a $\fr{\pi}{4}$ rotation and scaling,
\begin{equation}
\label{rot2}
\lb \begin{array}{c}
\om_L^3 \\ \om_R^3
\end{array} \rb
=
\fr{1}{\sqrt{2}}
\lb \begin{array}{cc}
\fr{1}{\sqrt{2}} & \fr{1}{\sqrt{2}} \\
\fr{-1}{\sqrt{2}} & \fr{1}{\sqrt{2}}
\end{array} \rb
\lb \begin{array}{c}
\fr{1}{i} \om_T^3 \\ \om_S^3
\end{array} \rb
=
\lb \begin{array}{c}
\fr{1}{2}(\fr{1}{i} \om_T^3 + \om_S^3) \\
\fr{1}{2}(-\fr{1}{i} \om_T^3 + \om_S^3)
\end{array} \rb
\end{equation}

Unlike other standard model roots, the roots of $so(3,1)$ are not all imaginary --- the coordinates along the $\om^3_T$ axis are real. The $Spin^+(3,1)$ Lie group of gravity, with Lie algebra $so(3,1)$, is neither simple nor compact --- it is isomorphic to $SL(2,\mathbb{C}) = SL(2,\mathbb{R}) \times SL(2,\mathbb{R})$. According to the $ADE$ classification of Lie groups it is still labeled $D2$ --- the same as $Spin(4)=SU(2) \times SU(2)$ --- since it has the same root system, albeit with one real axis.
 
\subsubsection{Electroweak $D2$}

The electroweak gauge field, $\f{W} \in \f{su}(2)_L$, acts on left-chiral doublets, such as $[ \ud{u}{}_L, \ud{d}{}_L ]$. The Pati-Salam GUT introduces a partner to this field, $\f{B}{}_1 \in \f{su}(2)_R$, acting on all right-chiral fermion doublets. Part of this field, $\f{B}{}_1^3 \fr{i}{2} \si_3 \in \f{u}(1)_R$, joins with the $u(1)_{B-L}$ complement, $\f{B}{}_2$, of the strong $su(3)$ to give the electroweak $\f{B} \in \f{u}(1)_Y$. The left-right electroweak partner fields may be joined in a $d2$ partner to gravity,
$$
so(4) = su(2)_L + su(2)_R
$$
Since both $W$ and $B_1$ act on the Higgs doublet, $[ \ph_+, \ph_0 ]$, it is sensible to consider the $4$ real fields of this Higgs doublet to be components of a vector acted on by the $so(4)$. This suggests we proceed as we did for gravity, using a complementary chiral matrix representation for the four orthonormal basis vectors of $Cl(4)$,
$$
\ga'_1 = \si_1 \otimes \si_1 \qquad
\ga'_2 = \si_1 \otimes \si_2 \qquad
\ga'_3 = \si_1 \otimes \si_3 \qquad
\ga'_4 = \si_2 \otimes 1
$$
These allow the Higgs vector field to be written as
$$
\ph = \ph^\mu \ga'_\mu
=
\lb \begin{array}{cc}
0 & -i \ph^4 \!+\! \ph^\va \si_\va \\
i \ph^4 \!+\! \ph^\va \si_\va & 0
\end{array} \rb
=
\lb \begin{array}{cccc}
0 & 0 & -\ph_1 & \ph_+ \\
0 & 0 & \p{-}\ph_- & \ph_0 \\
-\ph_0 & \ph_+ & 0 & 0 \\
\p{-}\ph_- & \ph_1 & 0 & 0
\end{array} \rb
\; \in \,
Cl^1(4)
$$
with coefficients equal to those of the Higgs doublet,
$$
\begin{array}{rclcrcl}
\ph_+ &=& \ph^1 - i \ph^2  &\qquad& \ph_- &=& \ph^1 + i \ph^2 \\
\ph_0 &=& - \ph^3 - i \ph^4  &\qquad& \ph_1 &=& - \ph^3 + i \ph^4
\end{array}
$$
The $d2 = so(4) = Cl^2(4)$ valued electroweak connection breaks up into two $su(2)$ parts,
$$
\f{w}_{ew} = \ha \f{w}_{ew}^{\mu \nu} \ga'_{\mu \nu}
=
\lb \begin{array}{cc}
( \f{V}{}^\ta \!+\! \f{U}{}^\ta ) \fr{i}{2} \si_\ta & 0 \\
0 & ( \f{V}{}^\ta \!-\! \f{U}{}^\ta ) \fr{i}{2} \si_\ta
\end{array} \rb
=
\lb \begin{array}{cc}
 \f{W}{}^\ta \fr{i}{2} \si_\ta & 0 \\
0 & \f{B}{}_1^\ta \fr{i}{2} \si_\ta
\end{array} \rb
$$
The $\f{U}$ and $\f{V}$ fields are analogous to the $\fr{1}{i} \f{\om}{}_T$ and $\f{\om}{}_S$ of gravity, and are related to the electroweak $\f{W}$ and $\f{B}{}_1$, analogous to the $\f{\om}{}_L$ and $\f{\om}{}_R$, by the same $\fr{\pi}{4}$ rotation and scaling (\ref{rot2}). The Cartan subalgebra,
\begin{eqnarray*}
C &=& \fr{1}{4} (W^3 + B_1^3) \ga'_{12} + \fr{1}{4} (W^3 - B_1^3) \ga'_{34} =
\frac{i}{2}
\lb \begin{array}{cccc}
W^3 & 0 & 0 & 0 \\
0 & -W^3 & 0 & 0 \\
0 & 0 & B_1^3 & 0 \\
0 & 0 & 0 & -B_1^3 \\
\end{array} \rb
\end{eqnarray*}

\TABLE[r]{
\begin{tabular}{@{\vrule width1.0pt}c@{\vrule width0.2pt}c@{\vrule width1.0pt}c@{\vrule width0.2pt}c@{\vrule width1.0pt}c@{\vrule width1.0pt}c@{\vrule width1.0pt}c@{\vrule width1.0pt}}\noalign{\hrule height 1.0pt}
\multicolumn{2}{@{\vrule width1.0pt}c@{\vrule width1.0pt}}{$D2_{ew}$} & $\, W^3 \,$ & $\, B_1^3 \,$ & $\, \fr{\sqrt{2}}{\sqrt{3}} B_2 \,$ & $\, \ha Y \,$ & $\, Q \,$ \\
\noalign{\hrule height 1pt}
\lyell{\mcir} & $\,W^+ \,$ & $1$ & $0$ & $0$ & $0$ & $1$ \\
\lyell{\mcir} & $W^-$ & $-1$ & $0$ & $0$ & $0$ & $-1$ \\
\noalign{\hrule height 0.2pt}
\white{\mcir} & $B_1^+$ & $0$ & $1$ & $0$ & $1$ & $1$ \\
\white{\mcir} & $B_1^-$ & $0$ & $-1$ & $0$ & $-1$ & $-1$ \\
\noalign{\hrule height 1.0pt}
\dyell{\msqu} & $\ph_+$ & $\nha$ & $\nfr{1}{2}$ & $0$ & $\nha$ & $1$ \\
\myell{\mdia} & $\ph_-$ & $\, \nfr{-1}{2} \,$ & $\, \nfr{-1}{2} \,$ & $0$ & $\nfr{-1}{2}$ & $-1$ \\
\mgray{\msqu} & $\ph_0$ & $\nfr{-1}{2}$ & $\nha$ & $0$ & $\nha$ & $0$ \\
\dgray{\mdia} & $\ph_1$ & $\nfr{1}{2}$ & $\nfr{-1}{2}$ & $0$ & $\nfr{-1}{2}$ & $0$ \\
\noalign{\hrule height 1.0pt}
$\,$\dyell{\mtri}$\,$& $\nu_L$ & $\nha$ & $0$ & $\nfr{1}{2}$ & $\nfr{-1}{2}$ & $0$ \\
\myell{\mtri} & $e_L$ & $\nfr{-1}{2}$ & $0$ & $\nfr{1}{2}$ & $\nfr{-1}{2}$ & $-1$ \\
\mgray{\mtri} & $\nu_R$ & $0$ & $\nha$ & $\nfr{1}{2}$ & $0$ & $0$ \\
\lgray{\mtri} & $e_R$ & $0$ & $\nfr{-1}{2}$ & $\nfr{1}{2}$ & $-1$ & $-1$ \\
\noalign{\hrule height 1.0pt}
\trip{\drora{\stri}}{\dygre{\stri}}{\dbvio{\stri}} & $u_L$ & $\nha$ & $0$ & $\nfr{-1}{6}$ & $\nfr{1}{6}$ & $\nfr{2}{3}$ \\
\trip{\mrora{\stri}}{\mygre{\stri}}{\mbvio{\stri}} & $d_L$ & $\nfr{-1}{2}$ & $0$ & $\, \nfr{-1}{6} \,$ & $\nfr{1}{6}$ & $\, \nfr{-1}{3} \,$ \\
\trip{\dred{\stri}}{\dgree{\stri}}{\dblue{\stri}} & $u_R$ & $0$ & $\nha$ & $\nfr{-1}{6}$ & $\nfr{2}{3}$ & $\nfr{2}{3}$ \\
\trip{\mred{\stri}}{\mgree{\stri}}{\mblue{\stri}} & $d_R$ & $0$ & $\nfr{-1}{2}$ & $\nfr{-1}{6}$ & $\, \nfr{-1}{3} \,$ & $\nfr{-1}{3}$ \\
\noalign{\hrule height 1.0pt}
\end{tabular} $\!\!\!\!\!$
\caption{Weights for electroweak $D2$, for $B_2$ from Table \ref{g23dt}, and electroweak hypercharge and charge.\label{wd2t}}
}
\noindent
gives root vectors and roots for the electroweak fields, such as $W^\pm$, and weight vectors and weights for the Higgs, such as
$$
\lb C , \fr{1}{2} (-\ga'_3 \!+\! i \ga'_4 ) \rb = i \! \lp W^3 (\nfr{-1}{2}) \!+\!  B_1^3 (\nfr{1}{2}) \rp \fr{1}{2} (-\ga'_3 \!+\! i \ga'_4 )
$$
for $\ph_0$. The fermions are acted on in the standard $4$, equivalent to the independent $su(2)_L$ and $su(2)_R$ action on left and right-chiral Weyl doublets, such as $[u_L, d_L]$ and $[u_R, d_R]$. The electroweak $D2$ weights for various fields are shown in Table \ref{wd2t}. 

The two right-chiral gauge fields, $\f{B}{}_1^\pm$, are not part of the standard model. They are a necessary part of the Pati-Salam GUT, and presumably have large masses or some other mechanism breaking left-right symmetry and impeding their detection. As in the Pati-Salam GUT, the $B_2$ weights from Table \ref{g23dt} and the $B_1^3$ weights may be scaled and rotated ((\ref{su3rot}) and (\ref{rot2})) into two new coordinates, including the weak hypercharge,
$$
\ha Y = B_1^3 - \fr{\sqrt{2}}{\sqrt{3}} B_2
$$
This scaling implies a weak hypercharge coupling constant of $g_1= \sqrt{\nfr{3}{5}}$ and Weinberg angle satisfying $\sin^2 \th_W = \nfr{3}{8}$, typical of almost all grand unified theories. There is also a new quantum number partner to the hypercharge, $X$, corresponding to the positive combination of quantum numbers $B_1^3$ and $B_2$. The hypercharge may be scaled and rotated with $W^3$ to give the electric charge,
$$
Q = W^3 + \ha Y
$$
These weights, shown in Table \ref{wd2t}, are in agreement with the known standard model quantum numbers, and justify our use of the corresponding particle labels.

\subsubsection{Graviweak $D4$}

The electroweak $d2=so(4)$ and gravitational $d2=so(3,1)$ combine as commuting parts of a graviweak $d4 = so(7,1)$. The $4$ Higgs fields, $\ph$, a vector of the electroweak $so(4)$, combine with the $4$ gravitational $so(3,1)$ vectors of the frame, $\f{e}$, into $16$ bivector valued fields, $\f{e} \ph$, of the graviweak $D4$ gauge group. This combination is achieved by adding the weights of Table \ref{gd2t} with those of Table \ref{wd2t} to obtain the weights of $D4$ in four dimensions, as shown in Table \ref{gwd4t}. The weights of the fermions also add to give their $D4$ weights. 

The fermion weights correspond to the fundamental positive-chiral spinor representation space, $8_{S+}$, of $D4$. To construct this explicitly, we use Trayling's model,\cite{Tray} and combine our $Cl(3,1)$ and $Cl(4)$ basis generators into eight Clifford basis vector elements of $Cl(7,1)$, represented as $(16 \times 16)$ matrices,
$$
\begin{array}{rclcrcl}
\Ga_1 &=& \si_2 \otimes \si_3 \otimes 1 \otimes \si_1 & \qquad &
\Ga'_1 &=& \si_2 \otimes \si_1 \otimes \si_1 \otimes 1 \\
\Ga_2 &=& \si_2 \otimes \si_3 \otimes 1 \otimes \si_2 & \qquad &
\Ga'_2 &=& \si_2 \otimes \si_1 \otimes \si_2 \otimes 1 \\
\Ga_3 &=& \si_2 \otimes \si_3 \otimes 1 \otimes \si_3 & \qquad &
\Ga'_3 &=& \si_2 \otimes \si_1 \otimes \si_3 \otimes 1 \\
\Ga_4 &=& i \si_1 \otimes 1 \otimes 1 \otimes 1 & \qquad &
\Ga'_4 &=& \si_2 \otimes \si_2 \otimes 1 \otimes 1
\end{array}
$$
\TABLE[r]{
\begin{tabular}{@{\vrule width1.0pt}c@{\vrule width0.2pt}c@{\vrule width1.0pt}c@{\vrule width0.2pt}c@{\vrule width0.2pt}c@{\vrule width0.2pt}c@{\vrule width1.0pt}}\noalign{\hrule height 1.0pt}
\multicolumn{2}{@{\vrule width1.0pt}c@{\vrule width1.0pt}}{$\,D4\,$} & $\, \ha \om_L^3 \,$ & $\, \ha \om_R^3 \,$ & $\, W^3 \,$ & $\, B_1^3 \,$ \\
\noalign{\hrule height 1pt}
\lgree{\mcir} & $\, \om_L^{\wedge/\vee} \,$ & $\pm 1$ & $0$ & $0$ & $0$ \\
\lgree{\mcir} & $\om_R^{\wedge/\vee}$ & $0$ & $\pm 1$ & $0$ & $0$ \\
\noalign{\hrule height 0.2pt}
\lyell{\mcir} & $\,W^\pm \,$ & $0$ & $0$ & $\pm 1$ & $0$ \\
\noalign{\hrule height 0.2pt}
\white{\mcir} & $B_1^\pm$ & $0$ & $0$ & $0$ & $\pm 1$ \\
\noalign{\hrule height 0.2pt}
\dyell{\msqu} & $e_T^{\wedge/\vee} \ph_+$ & $\nfr{\mp 1}{2}$ & $\nfr{\pm 1}{2}$ & $\nfr{1}{2}$ & $\nfr{1}{2}$ \\
\myell{\mdia} & $ \, e_T^{\wedge/\vee} \ph_- \,$ & $\,\nfr{\mp 1}{2}\,$ & $\,\nfr{\pm 1}{2}\,$ & $\,\nfr{-1}{2}\,$ & $\,\nfr{-1}{2}\,$ \\
\mrora{\msqu} & $e_T^{\wedge/\vee} \ph_0$ & $\nfr{\mp 1}{2}$ & $\nfr{\pm 1}{2}$ & $\nfr{-1}{2}$ & $\nha$ \\
\drora{\mdia} & $e_T^{\wedge/\vee} \ph_1$ & $\nfr{\mp 1}{2}$ & $\nfr{\pm 1}{2}$ & $\nfr{1}{2}$ & $\nfr{-1}{2}$ \\
\noalign{\hrule height 0.2pt}
\dyell{\msqu} & $e_S^{\wedge/\vee} \ph_+$ & $\nfr{\pm 1}{2}$ & $\nfr{\pm 1}{2}$ & $\nfr{1}{2}$ & $\nfr{1}{2}$ \\
\myell{\mdia} & $e_S^{\wedge/\vee} \ph_-$ & $\nfr{\pm 1}{2}$ & $\nfr{\pm 1}{2}$ & $\nfr{-1}{2}$ & $\nfr{-1}{2}$ \\
\mrora{\msqu} & $e_S^{\wedge/\vee} \ph_0$ & $\nfr{\pm 1}{2}$ & $\nfr{\pm 1}{2}$ & $\nfr{-1}{2}$ & $\nha$ \\
\drora{\mdia} & $e_S^{\wedge/\vee} \ph_1$ & $\nfr{\pm 1}{2}$ & $\nfr{\pm 1}{2}$ & $\nfr{1}{2}$ & $\nfr{-1}{2}$ \\
\noalign{\hrule height 1.0pt}
$\,\,$\dyell{\btri}$\,\,$ & $\nu_{eL}^{\wedge/\vee}$ & $\nfr{\pm 1}{2}$ & $0$ & $\nha$ & $0$ \\
\lyell{\btri} & $e_L^{\wedge/\vee}$ & $\nfr{\pm 1}{2}$ & $0$  & $\nfr{-1}{2}$ & $0$ \\
\dgray{\btri} & $\nu_{eR}^{\wedge/\vee}$ & $0$ & $\nfr{\pm 1}{2}$ & $0$ & $\nha$ \\
\lgray{\btri} & $e_R^{\wedge/\vee}$ & $0$ & $\nfr{\pm 1}{2}$  & $0$ & $\nfr{-1}{2}$ \\
\noalign{\hrule height 1.0pt}\end{tabular} $\!\!\!\!\!$
\caption{Graviweak $D4$ roots for $24$ bosons and weights for $8_{S+}$ fermions. \label{gwd4t}}}
\noindent
These allow us to build the spin connection, $\f{\om} = \ha \f{\om}{}^{\mu \nu} \Ga_{\mu \nu}$, the electroweak connection, $\f{w}{}_{ew} = \ha \f{\om}{}{}_{ew}^{\mu \nu} \Ga'_{\mu \nu}$, the frame, $\f{e} = \f{e}{}^\mu \Ga_\mu$, and the Higgs, $\ph = \ph^\mu \Ga'_\mu$, as $Cl(7,1)$ valued fields, with the same coefficients as before. The frame and Higgs multiply to give the frame-Higgs,
$\f{e} \ph = \f{e}{}^\mu \ph^\nu \Ga_\mu \Ga'_\nu$, a Clifford bivector valued 1-form. Together, these fields may be written as parts of a $Cl^2(7,1) = so(7,1)$ graviweak connection,
\begin{equation}
\f{H}{}_1 = \ha \f{\om} + \fr{1}{4} \f{e} \ph + \f{w}{}_{ew}
\label{H1}
\end{equation}
Since our chosen $Cl(7,1)$ representation is chiral, $\f{H}{}_1$ may be represented by its positive-chiral part, the $(8 \times 8)$ first quadrant of the $(16 \times 16)$ rep, shown here acting on a positive-chiral spinor, $8_{S+}$ :
$$
\lb \begin{array}{cccc}
\! \fr{1}{2} \om_L \!+\! \fr{i}{2} W^3 \! & W^+ & - \! \fr{1}{4} e_R \ph_1 & \fr{1}{4} e_R \ph_+ \\
W^- & \! \fr{1}{2} \om_L \!-\! \fr{i}{2} W^3 \! & \p{-} \fr{1}{4} e_R \ph_- & \fr{1}{4} e_R \ph_0 \\
-\fr{1}{4} e_L \ph_0 & \fr{1}{4} e_L \ph_+ & \! \fr{1}{2} \om_R \!+\! \fr{i}{2} B_1^3 \! & B_1^+ \\
\p{-}\fr{1}{4} e_L \ph_- & \fr{1}{4} e_L \ph_1 & B_1^- & \! \fr{1}{2} \om_R \!-\! \fr{i}{2} B_1^3 \!
\end{array} \rb
\lb \begin{array}{c}
\nu_{eL} \\ e_L \\ \nu_{eR} \\ e_R
\end{array}\rb
$$
The fractions, $\ha$ and $\fr{1}{4}$, multiplying fields in $\f{H}{}_1$ are necessary for fitting gravity and the electroweak connection together in $D4$, and for obtaining the correct dynamics from curvature.

The first $24$ weights in Table \ref{gwd4t} are the roots of $D4$. This particular root system has a uniquely beautiful set of symmetries called \textbf{triality},\cite{Baez} rotations of the root system by $\fr{2 \pi}{3}$ that leave it invariant. A triality rotation matrix, $T$, can permute the coordinates of the root system,
$$
\lb \begin{array}{c}
\ha {\om'}^3_L \\ \ha {\om'}^3_R \\ {W'}^3 \\ {B'}_1^3
\end{array} \rb
=
\lb \begin{array}{cccc}
0 & 0 & 0 & 1 \\
1 & 0 & 0 & 0 \\
0 & 0 & 1 & 0 \\
0 & 1 & 0 & 0
\end{array} \rb
\lb \begin{array}{c}
\ha \om^3_L \\ \ha \om^3_R \\ W^3 \\ B_1^3
\end{array} \rb
=
\lb \begin{array}{c}
B_1^3 \\ \fr{1}{2} \om^3_L \\ W^3 \\ \ha \om^3_R  
\end{array} \rb
$$
taking each root to its first triality partner, then to its second, and back --- satisfying $\,T^3=1$. As an example, the above triality rotation gives
$$
T T T \om_R^\wedge = T T B_1^+ = T \om_L^\wedge = \om_R^\wedge
$$
showing the equivalence of these roots under this triality rotation. Six of the roots,
$$
\{
W^+, \,
W^- , \,
e_S^\wedge\ph_+, \,
e_S^\wedge \ph_0, \,
e_S^\vee \ph_-, \,
e_S^\vee \ph_1
\}
$$
are their own triality partners --- they lie in the plane orthogonal to this triality rotation.

The last $8$ weights of Table \ref{gwd4t}, representing one generation of leptons as $8_{S+}$, are rotated by triality into the other fundamental representation spaces of $D4$: the negative-chiral spinor and the vector,
$$
T \, 8_{S+} = 8_{S-} \qquad T \, 8_{S-} = 8_V  \qquad T \, 8_V = 8_{S+}
$$
These two new sets of weights are equivalent to the $8_{S+}$ under this triality rotation --- they carry the same quantum numbers and have the same interactions with the triality rotated roots of $D4$. Given this relationship, we tentatively consider these three triality partners of $8_{S+}$ as the three generations of fermions, such as
$$
T e_L^\wedge = \mu_L^\wedge
\qquad
T \mu_L^\wedge = \ta_L^\wedge
\qquad
T \ta_L^\wedge = e_L^\wedge
$$
The complete set of weights, $D4 + (8_{S+} \!+ 8_{S-} \!+ 8_{V})$, including these new triality partners, is the root system of the rank four simple exceptional group, $F4$.

\subsubsection{$F4$}

\begin{floatingtable}
{\centerline{\parbox{.35\textwidth}{\centerline{
$\!\!\!\!\!\!\!\!\!\!\!\!\!\!\!\!\!\!\!\!\!\!\!\!\!\!\!\!\!\!\!\!\!\!\!\!\!\!\!\!\!\!\!\!\!\!\!\!\!\!\!\!\!\!\!\!\!\!\!\!\!\!\!\!\!\!\!\!\!\!\!\!\!\!\!\!\!$
\begin{tabular}{c}
\begin{tabular}
{@{\vrule width1.0pt}c@{\vrule width0.2pt}c@{\vrule width1.0pt}c@{\vrule width0.2pt}c@{\vrule width0.2pt}c@{\vrule width0.2pt}c@{\vrule width1.0pt}}
\noalign{\hrule height 1.0pt}
\multicolumn{2}{@{\vrule width1.0pt}c@{\vrule width1.0pt}}{$8_{S-}$} & $\, \ha \om_L^3 \,$ & $\, \ha \om_R^3 \,$ & $\, W^3 \,$ & $\, B_1^3 \,$ \\
\noalign{\hrule height 0.2pt}
\multicolumn{2}{@{\vrule width1.0pt}c@{\vrule width1.0pt}}{$\;\;\;\;$tri$\;\;\;\;$} & $\, B_1^3 \,$ & $\, \fr{1}{2} \om_L^3 \,$ & $\, W^3 \,$ & $\, \ha \om_R^3 \,$ \\
\noalign{\hrule height 1.0pt}
$\,$\dyell{\mtri}$\,$ & $\, \nu_{\mu L}^{\wedge/\vee} \,$ & $0$ & $\nfr{\pm 1}{2}$ & $\nha$ & $0$ \\
\lyell{\mtri} & $\mu_L^{\wedge/\vee}$ & $0$ & $\nfr{\pm 1}{2}$ & $\nfr{-1}{2}$ & $0$ \\
\dgray{\mtri} & $\nu_{\mu R}^{\wedge/\vee}$ & $\nfr{1}{2}$ & $0$ & $0$ & $\nfr{\pm 1}{2}$ \\
\lgray{\mtri} & $\mu_R^{\wedge/\vee}$ & $\nfr{-1}{2}$ & $0$ & $0$ & $\nfr{\pm 1}{2}$ \\
\noalign{\hrule height 1.0pt}
\end{tabular}
\\
\\
\\
\\
\begin{tabular}
{@{\vrule width1.0pt}c@{\vrule width0.2pt}c@{\vrule width1.0pt}c@{\vrule width0.2pt}c@{\vrule width0.2pt}c@{\vrule width0.2pt}c@{\vrule width1.0pt}}
\noalign{\hrule height 1.0pt}
\multicolumn{2}{@{\vrule width1.0pt}c@{\vrule width1.0pt}}{$8_{V}$} & $\, \ha \om_L^3 \,$ & $\, \ha \om_R^3 \,$ & $\, W^3 \,$ & $\, B_1^3 \,$ \\
\noalign{\hrule height 0.2pt}
\multicolumn{2}{@{\vrule width1.0pt}c@{\vrule width1.0pt}}{$\;\;\;\;$tri$\;\;\;\;$} & $\, \ha \om_R^3 \,$ & $\, B_1^3 \,$ & $\, W^3 \,$ & $\, \fr{1}{2} \om_L^3 \,$ \\
\noalign{\hrule height 1.0pt}
$\,$\dyell{\stri}$\,$ & $\, \nu_{\ta L}^{\wedge/\vee} \,$ & $0$ & $0$ & $\nfr{1}{2}$ & $\nfr{\pm 1}{2}$ \\
\lyell{\stri} & $\ta_L^{\wedge/\vee}$ & $0$ & $0$ & $\nfr{-1}{2}$ & $\nfr{\pm 1}{2}$ \\
\dgray{\stri} & $\nu_{\ta R}^{\wedge/\vee}$ & $\nfr{\pm 1}{2}$ & $\nfr{1}{2}$ & $0$ & $0$ \\
\lgray{\stri} & $\ta_R^{\wedge/\vee}$ & $\nfr{\pm 1}{2}$ & $\nfr{-1}{2}$ & $0$ & $0$ \\
\noalign{\hrule height 1.0pt}
\end{tabular}
\end{tabular}
}}
\parbox{.35\textwidth}{\centerline{
\epsfig{file=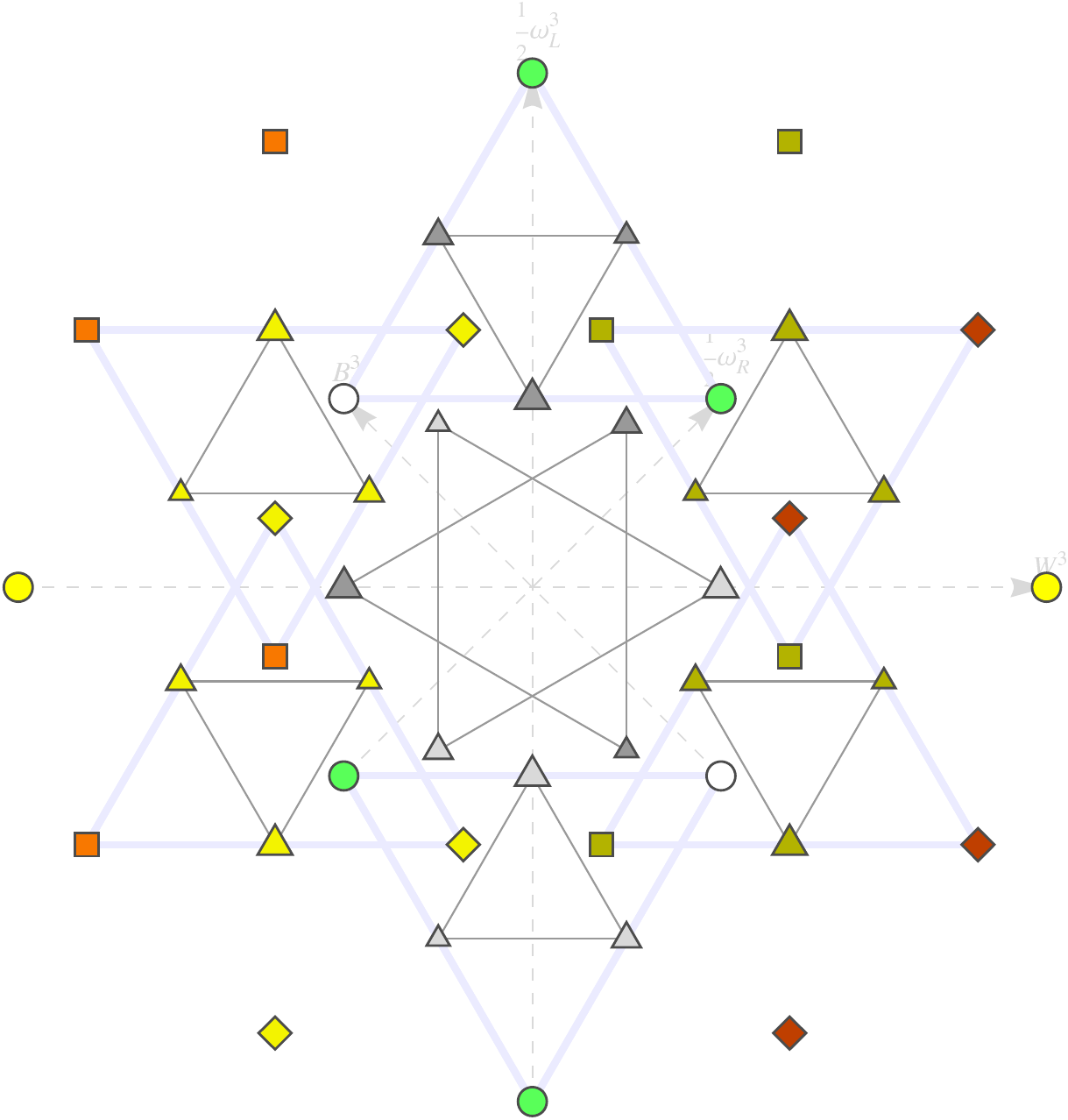, width=4in}}}}
\caption{The $8_{S-}$ and $8_{V}$ weights complete the $D4$ and $8_{S+}$ weight system of Table \protect\ref{gwd4t} to form the $F4$ root system. The $48$ roots are projected from four dimensions to two and plotted, with lines shown between triality partners.\label{f4t}}}
\end{floatingtable}
\noindent
The $48$ roots of $F4$ are shown in Tables \ref{gwd4t} and \ref{f4t}. These roots, in four dimensions, are the vertices of the $24$-cell polytope and its dual. Using the breakdown of $F4$ into $D4$ and the three triality-equivalent fundamental representation spaces,
\begin{equation}
f4 = d4 + (8_{S+}\!+ 8_{S-}\!+ 8_{V}) = so(7,1) + (8\!+\!8\!+\!8)
\label{generations}
\end{equation}
the graviweak bosons (\ref{H1}) and three generations of leptons (or quarks) may be written as parts of a $F4$ connection,
$$
\f{H}{}_1 + ( \ud{\nu}{}_e + \ud{e} ) + ( \ud{\nu}{}_\mu + \ud{\mu}) + (\ud{\nu}{}_\ta + \ud{\ta}) 
$$
Although we are labeling triality partners as fermions of different generations, the exact relationship between triality and generations is more complicated and not yet clear to the author. One clue is that the triality partners of $F4$ (connected in the figure by pale blue and thin gray lines) may be collapsed to their midpoints to get a $g2$ subalgebra,
$$
\fr{1}{3}(1 + T + T^2) f4 = g2 \subset f4
$$
This triality collapse might relate to a description of graviweak interactions with a group smaller than $F4.$\cite{Nest,Alex} It also suggests physical fermions may be linear combinations of triality parners, such as
$$
\mu_L^\wedge = a \, f_L^\wedge + b \, T f_L^\wedge + c \, T T f_L^\wedge   
$$
Guided by this triality symmetry, we will continue to label the triality partners with generation labels --- though this should be understood as an idealization of a more complex and as yet unclear relationship between physical particles and triality partners.

If we wished, we could write the constituent particles of $F4$ as matrix elements of its smallest irreducible, $26$ dimensional representation,\cite{Bern} as we did for the gluons and quarks (\ref{7rep}) in $G2$. We can also compute particle interactions by adding the roots in Tables \ref{gwd4t} and \ref{f4t}, such as
$$
e_L^\wedge + e_T^\wedge \ph_+ = \nu_{eR}^\wedge
$$
between an electron, a frame-Higgs, and an electron neutrino. These graviweak interactions, described by the structure of $F4$, do not involve anti-fermions or color; to include all standard model interactions we will have to combine $F4$ and $G2$.

\subsection{$F4$ and $G2$ together}

The coordinate axes chosen in Tables \ref{gwd4t} and \ref{f4t} are a good choice for expressing the quantum numbers for gravity and the electroweak fields, but they are not the standard axes for describing the $F4$ root system. We can rotate to our other coordinate system,
$$
\{ \fr{1}{2i} \om_T^3, \fr{1}{2} \om_S^3 , U^3, V^3 \}
$$
using a pair of $\fr{\pi}{4}$ rotations (\ref{rot2}) and thereby express the $48$ roots of $F4$ in standard coordinates, shown in Table \ref{f4andg2}. These coordinate values are described by various permutations of $\pm 1$, $\pm \nha$, and $0$; and a similar description of the $G2$ and $U(1)$ weights from Table \ref{g23dt} is also presented.
\TABLE{
{\begin{tabular}{cc}
\begin{tabular}
{@{\vrule width1.0pt}c@{\vrule width0.2pt}c@{\vrule width1.0pt}c@{\vrule width0.4pt}c@
{\vrule width0.4pt}c@{\vrule width0.4pt}c@{\vrule width1.0pt}l@{\vrule width1.0pt}c@{\vrule width1.0pt}}
\noalign{\hrule height 1.0pt}
\multicolumn{2}{@{\vrule width1.0pt}c@{\vrule width1.0pt}}{$F4$} & $\, \fr{1}{2i} \om_T^3 \,$ & $\, \ha \om_S^3 \,$ & $\, U^3 \,$ & $\; V^3 \;$ & $\,$perms$\,$ & \# \\
\noalign{\hrule height 1pt}
\lgray{\mcir} & $\, so(7,1) \,$ & \multicolumn{4}{c@{\vrule width1.0pt}}{$\!\pm 1 \; \pm \! 1 $} & $\,$all & $\, 24 \,$ \\
\noalign{\hrule height .5pt}
\myell{\btri} & $\, 8_{S+} \,$ & \multicolumn{1}{c@{\vrule width0.0pt}}{$\!\!\nfr{\pm 1}{2} \,$} & \multicolumn{1}{c@{\vrule width0.0pt}}{$\!\!\nfr{\pm 1}{2} \,$} & \multicolumn{1}{c@{\vrule width0.0pt}}{$\!\!\nfr{\pm 1}{2} \,$} & \multicolumn{1}{c@{\vrule width1.0pt}}{$\!\!\nfr{\pm 1}{2} \,$} & $\,$even\# $>0\,$ & $\, 8 \,$ \\
\noalign{\hrule height 0.2pt}
$\,$\myell{\mtri}$\,$& $\, 8_{S-} \,$ & \multicolumn{1}{c@{\vrule width0.0pt}}{$\!\!\nfr{\pm 1}{2} \,$} & \multicolumn{1}{c@{\vrule width0.0pt}}{$\!\!\nfr{\pm 1}{2} \,$} & \multicolumn{1}{c@{\vrule width0.0pt}}{$\!\!\nfr{\pm 1}{2} \,$} & \multicolumn{1}{c@{\vrule width1.0pt}}{$\!\!\nfr{\pm 1}{2} \,$} & $\,$odd\# $>0\,$ & $\, 8 \,$ \\
\noalign{\hrule height 0.2pt}
\myell{\stri} & $\, 8_{V} \,$ & \multicolumn{4}{c@{\vrule width1.0pt}}{$\!\pm 1 \,$} & $\,$all & $\, 8 \,$ \\
\noalign{\hrule height 1.0pt}
\end{tabular}
$\;$ &
\begin{tabular}
{@{\vrule width1.0pt}c@{\vrule width0.2pt}c@{\vrule width1.0pt}c@{\vrule width0.4pt}c@
{\vrule width0.4pt}c@{\vrule width1.0pt}l@{\vrule width1.0pt}c@{\vrule width1.0pt}}
\noalign{\hrule height 1.0pt}
\multicolumn{2}{@{\vrule width1.0pt}c@{\vrule width1.0pt}}{$\, G2+U(1) \,$} & $\, x \,$ & $\, y \,$ & $\; z \;$ & $\,$perms$\,$ & $\,$\#$\,$ \\
\noalign{\hrule height 1pt}
\lblue{\mcir} & $\, su(3) \,$ & \multicolumn{3}{c@{\vrule width1.0pt}}{$\!1\;\;-\!\!1$} & $\,$all & $\, 6 \,$ \\
\noalign{\hrule height .5pt}
\trip{\mred{\stri}}{\mgree{\stri}}{\mblue{\stri}} & $\, q_I \,$ & \multicolumn{1}{c@{\vrule width0.0pt}}{$\!\!\nfr{\pm 1}{2} \,$} & \multicolumn{1}{c@{\vrule width0.0pt}}{$\!\!\nfr{\pm 1}{2} \,$} & \multicolumn{1}{c@{\vrule width1.0pt}}{$\!\!\nfr{\pm 1}{2} \,$} & $\,$two $>0\,$ & $\, 3 \,$ \\
\noalign{\hrule height 0.2pt}
\trip{\mred{\sutr}}{\mgree{\sutr}}{\mblue{\sutr}} & $\, \bar{q}_I \,$ & \multicolumn{1}{c@{\vrule width0.0pt}}{$\!\!\nfr{\pm 1}{2} \,$} & \multicolumn{1}{c@{\vrule width0.0pt}}{$\!\!\nfr{\pm 1}{2} \,$} & \multicolumn{1}{c@{\vrule width1.0pt}}{$\!\!\nfr{\pm 1}{2} \,$} & $\,$one $>0\,$ & $\, \bar{3} \,$ \\
\noalign{\hrule height .5pt}
$\,$\mgray{\mtri}$\,$ & $\, l \,$ & $\nfr{-1}{2}$ & $\nfr{-1}{2}$ & $\nfr{-1}{2}$ & $\,$one & $\, 1 \,$ \\
\noalign{\hrule height 0.2pt}
$\,$\mgray{\mutr}$\,$ & $\, \bar{l} \,$ & $\nfr{1}{2}$ & $\nfr{1}{2}$ & $\nfr{1}{2}$ & $\,$one & $\, \bar{1} \,$ \\
\noalign{\hrule height .5pt}
\trip{\mred{\stri}}{\mgree{\stri}}{\mblue{\stri}} & $\, q_{II} \,$ & \multicolumn{3}{c@{\vrule width1.0pt}}{$\!-\!1$} & $\,$all & $\, 3 \,$ \\
\noalign{\hrule height 0.2pt}
$\,$\trip{\mred{\sutr}}{\mgree{\sutr}}{\mblue{\sutr}}$\,$ & $\, \bar{q}_{II} \,$ & \multicolumn{3}{c@{\vrule width1.0pt}}{$\!1$} & $\,$all & $\, \bar{3} \,$ \\
\noalign{\hrule height .5pt}
\trip{\mred{\msqu}}{\mgree{\msqu}}{\mblue{\msqu}} & $\, q_{III} \,$ & \multicolumn{3}{c@{\vrule width1.0pt}}{$\!1\;\;\;\;1$} & $\,$all & $\, 3 \,$ \\
\noalign{\hrule height 0.2pt}
\trip{\mred{\mdia}}{\mgree{\mdia}}{\mblue{\mdia}} & $\, \bar{q}_{III} \,$ & \multicolumn{3}{c@{\vrule width1.0pt}}{$\!\!-\!1\;-\!\!1$} & $\,$all & $\, \bar{3} \,$ \\
\noalign{\hrule height 1.0pt}
\end{tabular}
\\
\end{tabular}}
\caption{Roots of $F4$ and weights of Table \protect\ref{g23dt} described with allowed permutations of coordinate values.\label{f4andg2}}
}

To completely describe every field in the standard model and gravity we need to combine these two sets of quantum numbers. The graviweak $F4$ root system includes the two quantum numbers of $so(3,1)$ gravity and the two of the $su(2)_L$ and $su(2)_R$ electroweak fields, with three generations of fermions related through $so(7,1)$ triality (\ref{generations}). The $G2$ weight system includes the three quantum numbers of the $su(3)$ strong fields and a $u(1)_{B-L}$ contributing to hypercharge, with fermions and anti-fermions related through duality (\ref{su31}). To match the quantum numbers of all known standard model and gravitational fields, the $so(7,1)$ of $F4$ and $su(3)$ and $u(1)_{B-L}$ of $G2+U(1)$ must act on three generations of $8$ fermions for each of the $3$ colors of quark, $1$ uncolored lepton, and their anti-particles,  
\begin{equation}
\label{smalgebra}
so(7,1) + (su(3) + u(1)) + (8 + 8 + 8)\times(3+\bar{3}+1+\bar{1})
\end{equation}
as depicted in the periodic table, Figure \ref{ptsm}. The weights of these $222$ elements --- corresponding to the quantum numbers of all gravitational and standard model fields --- exactly match $222$ roots out of the $240$ of the largest simple exceptional Lie group, $E8$.  

\pagebreak

\subsection{$E8$}
\TABLE[r]{
\begin{tabular}
{@{\vrule width1.0pt}c@{\vrule width1.0pt}c@{\vrule width0.2pt}c@{\vrule width0.2pt}c@{\vrule width0.2pt}c@{\vrule width0.2pt}c@{\vrule width0.2pt}c@{\vrule width0.2pt}c@{\vrule width0.2pt}c@{\vrule width1.0pt}c@{\vrule width1.0pt}}
\noalign{\hrule height 1.0pt}
$\, E8 \,$ & $\, x^1 \,$ & $\, x^2 \,$ & $\, x^3 \,$ & $\, x^4 \,$ & $\, x^5 \,$ & $\, x^6 \,$ & $\, x^7 \,$ & $\, x^8 \,$ & \#  \\
\noalign{\hrule height 1.0pt}
$\, so(16) \,$ & \multicolumn{8}{c@{\vrule width1.0pt}}{$ \pm 1 \; \pm \! 1 \,$ all perms$\,$} & $\, 112 \,$  \\
\noalign{\hrule height 0.5pt}
$\, 16_{S+} \,$ & \multicolumn{8}{c@{\vrule width1.0pt}}{$\nfr{\pm 1}{2} \; ... \,$ even\# $>0$} & $128$  \\
\noalign{\hrule height 1pt}
\end{tabular}
\caption{The $240$ roots of $E8$.\label{e8tsmall}}}
$\,$

\noindent
{\it``$E8$ is perhaps the most beautiful structure \qquad \\
in all of mathematics, but it's very complex.'' \\
--- Hermann Nicolai}

$\vp{\Big(_{\Big(}}$

\noindent
Just as we joined the weights of $D2_G$ and $D2_{ew}$ to form the $F4$ graviweak root system, the weights of $F4$ and $G2$ may be joined to form the roots of $E8$ --- the vertices of the $E8$ polytope --- shown in Table \ref{e8tsmall}. Combining these weights in eight dimensions requires the introduction of a new quantum number, $w$, with values determined by the $F4$ and $G2$ numbers. These quantum numbers uniquely identify each root of $E8$ as an elementary particle --- Table \ref{e8tmedium}.  $\vp{\big(_{\Big(}}$

\TABLE{
\begin{tabular}
{@{\vrule width1.0pt}c@{\vrule width0.2pt}c@{\vrule width1.0pt}c@{\vrule width0.4pt}c@
{\vrule width0.4pt}c@{\vrule width0.4pt}c@{\vrule width0.4pt}c@{\vrule width0.4pt}c@{\vrule width0.4pt}c@{\vrule width0.4pt}c@{\vrule width1.0pt}c@{\vrule width1.0pt}c@{\vrule width1.0pt}c@{\vrule width1.0pt}}
\noalign{\hrule height 1.0pt}
\multicolumn{2}{@{\vrule width1.0pt}c@{\vrule width1.0pt}}{$E8$} & $\, \fr{1}{2i} \om_T^3 \,$ & $\, \ha \om_S^3 \,$ & $\, U^3 \,$ & $\; V^3 \;$ & $\; w \;$ & $\;\;\; x \;\;\;$ & $\;\;\; y \;\;\;$ & $\;\;\; z \;\;\;$ & $\; F4 \;$ & $\; G2 \;$ & \# \\
\noalign{\hrule height 1pt}
$\,$\lgree{\mcir} \lgree{\mcir}$\,$ & $\; \om_{L}^{\wedge/\vee} \;\; \om_{R}^{\wedge/\vee} \;$ & \multicolumn{1}{c@{\vrule width.0pt}}{$\pm 1$} & $\pm 1$ & \multicolumn{2}{c@{\vrule width.4pt}}{$0\;\;$} & $0$ & \multicolumn{3}{c@{\vrule width1.0pt}}{$0 \;\;$} & $D2_G$ & $1$ & $\, 4 \,$ \\
\noalign{\hrule height .2pt}
$\,$\lyell{\mcir} \white{\mcir}$\,$ & $\; W^\pm \;\; B_1^\pm \;$ & \multicolumn{2}{c@{\vrule width.4pt}}{$0\;\;$} &  \multicolumn{1}{c@{\vrule width.0pt}}{$\pm 1$} & $\pm 1$ & $0$ & \multicolumn{3}{c@{\vrule width1.0pt}}{$0 \;\;$} & $D2_{ew}$ & $1$ & $\, 4 \,$ \\
\noalign{\hrule height .2pt}
$\,$\dyell{\msqu}$\,$\myell{\mdia}$\,$\drora{\mdia}$\,$\mrora{\msqu}$\,$ & $\; e\ph_+ \;\; e\ph_- \;\; e\ph_1 \;\; e\ph_0 \;$ & \multicolumn{2}{c@{\vrule width.4pt}}{$\pm 1$} & \multicolumn{2}{c@{\vrule width.4pt}}{$\pm 1$} & $0$ & \multicolumn{3}{c@{\vrule width1.0pt}}{$0 \;\;$} & $\, 4 \times 4 \,$ & $1$ & $\, 16 \,$ \\
\noalign{\hrule height .5pt}
$\,$\dyell{\btri}\myell{\btri}\mgray{\btri}\lgray{\btri}$\,$ & $\; \nu_{eL} \;\; e_L \;\; \nu_{eR} \;\; e_R \;$ & \multicolumn{4}{c@{\vrule width.4pt}}{$\nfr{\pm 1}{2} \; ...$ even\#$>\!0\,$} & $\,\nfr{-1}{2}\,$ & $\,\nfr{-1}{2}\,$ & $\,\nfr{-1}{2}\,$ & $\,\nfr{-1}{2}\,$ & $\, 8_{S+} \,$ & $l$ & $\, 8 \,$ \\
\noalign{\hrule height .2pt}
$\,$\dyell{\butr}\myell{\butr}\mgray{\butr}\lgray{\butr}$\,$ & $\; \bar{\nu}_{eL} \;\; \bar{e}_L \;\; \bar{\nu}_{eR} \;\; \bar{e}_R \;$ & \multicolumn{4}{c@{\vrule width.4pt}}{$\nfr{\pm 1}{2} \; ...$ even\#$>\!0\,$} & $\,\nfr{1}{2}\,$ & $\,\nfr{1}{2}\,$ & $\,\nfr{1}{2}\,$ & $\,\nfr{1}{2}\,$ & $\, 8_{S+} \,$ & $\bar{l}$ & $\, 8 \,$ \\
\noalign{\hrule height .2pt}
$\,$\trip{\drora{\btri}}{\dygre{\btri}}{\dbvio{\btri}}\trip{\mrora{\btri}}{\mygre{\btri}}{\mbvio{\btri}}\trip{\dred{\btri}}{\dgree{\btri}}{\dblue{\btri}}\trip{\mred{\btri}}{\mgree{\btri}}{\mblue{\btri}}$\,$ & $\, u_L \;\; d_L \;\; u_R \;\; d_R \;$ & \multicolumn{4}{c@{\vrule width.4pt}}{$\nfr{\pm 1}{2} \; ...$ even\#$>\!0\,$} & $\,\nfr{-1}{2}\,$ & \multicolumn{3}{c@{\vrule width1pt}}{$\nfr{\pm 1}{2} \; ...$ two$>\!0\,$} & $\, 8_{S+} \,$ & $q_I$ & $\, 24 \,$ \\
\noalign{\hrule height .2pt}
$\,$\trip{\drora{\butr}}{\dygre{\butr}}{\dbvio{\butr}}\trip{\mrora{\butr}}{\mygre{\butr}}{\mbvio{\butr}}\trip{\dred{\butr}}{\dgree{\butr}}{\dblue{\butr}}\trip{\mred{\butr}}{\mgree{\butr}}{\mblue{\butr}}$\,$ & $\, \bar{u}_L \;\; \bar{d}_L \;\; \bar{u}_R \;\; \bar{d}_R \;$ & \multicolumn{4}{c@{\vrule width.4pt}}{$\nfr{\pm 1}{2} \; ...$ even\#$>\!0\,$} & $\,\nfr{1}{2}\,$ & \multicolumn{3}{c@{\vrule width1pt}}{$\nfr{\pm 1}{2} \; ...$ one$>\!0\,$} & $\, 8_{S+} \,$ & $\bar{q}_I$ & $\, 24 \,$ \\
\noalign{\hrule height .5pt}
$\,$\dyell{\mtri}\myell{\mtri}\mgray{\mtri}\lgray{\mtri}$\,$ & $\; \nu_{\mu L} \;\; \mu_L \;\; \nu_{\mu R} \;\; \mu_R \;$ & \multicolumn{4}{c@{\vrule width.4pt}}{$\nfr{\pm 1}{2} \; ...$ odd\#$>\!0\,$} & $\,\nfr{-1}{2}\,$ & $\,\nfr{1}{2}\,$ & $\,\nfr{1}{2}\,$ & $\,\nfr{1}{2}\,$ & $\, 8_{S-} \,$ & $l$ & $\, 8 \,$ \\
\noalign{\hrule height .2pt}
$\,$\dyell{\mutr}\myell{\mutr}\mgray{\mutr}\lgray{\mutr}$\,$ & $\; \bar{\nu}_{\mu L} \;\; \bar{\mu }_L \;\; \bar{\nu}_{\mu R} \;\; \bar{\mu }_R \;$ & \multicolumn{4}{c@{\vrule width.4pt}}{$\nfr{\pm 1}{2} \; ...$ odd\#$>\!0\,$} & $\,\nfr{1}{2}\,$ & $\,\nfr{-1}{2}\,$ & $\,\nfr{-1}{2}\,$ & $\,\nfr{-1}{2}\,$ & $\, 8_{S-} \,$ & $\bar{l}$ & $\, 8 \,$ \\
\noalign{\hrule height .2pt}
$\,$\trip{\drora{\mtri}}{\dygre{\mtri}}{\dbvio{\mtri}}\trip{\mrora{\mtri}}{\mygre{\mtri}}{\mbvio{\mtri}}\trip{\dred{\mtri}}{\dgree{\mtri}}{\dblue{\mtri}}\trip{\mred{\mtri}}{\mgree{\mtri}}{\mblue{\mtri}}$\,$ & $\, c_L \;\; s_L \;\; c_R \;\; s_R \;$ & \multicolumn{4}{c@{\vrule width.4pt}}{$\nfr{\pm 1}{2} \; ...$ odd\#$>\!0\,$} & $\,\nfr{1}{2}\,$ & \multicolumn{3}{c@{\vrule width1pt}}{$\nfr{\pm 1}{2} \; ...$ two$>\!0\,$} & $\, 8_{S-} \,$ & $q_I$ & $\, 24 \,$ \\
\noalign{\hrule height .2pt}
$\,$\trip{\drora{\mutr}}{\dygre{\mutr}}{\dbvio{\mutr}}\trip{\mrora{\mutr}}{\mygre{\mutr}}{\mbvio{\mutr}}\trip{\dred{\mutr}}{\dgree{\mutr}}{\dblue{\mutr}}\trip{\mred{\mutr}}{\mgree{\mutr}}{\mblue{\mutr}}$\,$ & $\, \bar{c}_L \;\; \bar{s}_L \;\; \bar{c}_R \;\; \bar{s}_R \;$ & \multicolumn{4}{c@{\vrule width.4pt}}{$\nfr{\pm 1}{2} \; ...$ odd\#$>\!0\,$} & $\,\nfr{-1}{2}\,$ & \multicolumn{3}{c@{\vrule width1pt}}{$\nfr{\pm 1}{2} \; ...$ one$>\!0\,$} & $\, 8_{S-} \,$ & $\bar{q}_I$ & $\, 24 \,$ \\
\noalign{\hrule height .5pt}
$\,$\dyell{\stri}\myell{\stri}\mgray{\stri}\lgray{\stri}$\,$ & $\; \nu _{\ta L} \;\; \ta_L \;\; \nu _{\ta R} \;\; \ta_R \;$ & \multicolumn{4}{c@{\vrule width.4pt}}{$\pm 1$} & $\, 1 \,$ & \multicolumn{3}{c@{\vrule width1.0pt}}{$0 \;\;$} & $\, 8_{V} \,$ & $1$ & $\, 8 \,$ \\
\noalign{\hrule height .2pt}
$\,$\dyell{\sutr}\myell{\sutr}\mgray{\sutr}\lgray{\sutr}$\,$ & $\; \bar{\nu }_{\ta L} \;\; \bar{\ta}_L \;\; \bar{\nu}_{\ta R} \;\; \bar{\ta}_R \;$ & \multicolumn{4}{c@{\vrule width.4pt}}{$\pm 1$} & $\, -\!1 \,$ & \multicolumn{3}{c@{\vrule width1.0pt}}{$0 \;\;$} & $\, 8_{V} \,$ & $1$ & $\, 8 \,$ \\
\noalign{\hrule height .2pt}
$\,$\trip{\drora{\stri}}{\dygre{\stri}}{\dbvio{\stri}}\trip{\mrora{\stri}}{\mygre{\stri}}{\mbvio{\stri}}\trip{\dred{\stri}}{\dgree{\stri}}{\dblue{\stri}}\trip{\mred{\stri}}{\mgree{\stri}}{\mblue{\stri}}$\,$ & $\, t_L \;\; b_L \;\; t_R \;\; b_R \;$ & \multicolumn{4}{c@{\vrule width.4pt}}{$\pm 1$} & $\,0 \,$ & \multicolumn{3}{c@{\vrule width1.0pt}}{$-\!1 \;\;$} & $\, 8_{V} \,$ & $q_{II}$ & $\, 24 \,$ \\
\noalign{\hrule height .2pt}
$\,$\trip{\drora{\sutr}}{\dygre{\sutr}}{\dbvio{\sutr}}\trip{\mrora{\sutr}}{\mygre{\sutr}}{\mbvio{\sutr}}\trip{\dred{\sutr}}{\dgree{\sutr}}{\dblue{\sutr}}\trip{\mred{\sutr}}{\mgree{\sutr}}{\mblue{\sutr}}$\,$ & $\, \bar{t}_L \;\; \bar{b}_L \;\; \bar{t}_R \;\; \bar{b}_R \;$ & \multicolumn{4}{c@{\vrule width.4pt}}{$\pm 1$} & $\,0 \,$ & \multicolumn{3}{c@{\vrule width1.0pt}}{$1 \;\;$} & $\, 8_{V} \,$ & $\bar{q}_{II}$ & $\, 24 \,$ \\
\noalign{\hrule height .5pt}
$\,$\lblue{\mcir}$\,$ & $\, g \;$ & \multicolumn{4}{c@{\vrule width.4pt}}{$0$} & $\,0 \,$ & \multicolumn{3}{c@{\vrule width1.0pt}}{$1 \;\; -\!\!1 \;\;$} & $1$ & $\, A2 \,$ & $\, 6 \,$ \\
\noalign{\hrule height .2pt}
$\,$\trip{\mred{\bsqu}}{\mgree{\bsqu}}{\mblue{\bsqu}} \trip{\mred{\bdia}}{\mgree{\bdia}}{\mblue{\bdia}}$\,$ & $\, x_1 \Ph \,$ & \multicolumn{4}{c@{\vrule width.4pt}}{$0$} & $\, -\!1 \,$ & \multicolumn{3}{c@{\vrule width1pt}}{$\pm 1 \;\;$} & $\, 1 \,$ & $q_{II}$ & $\, 6 \,$ \\
\noalign{\hrule height .2pt}
$\,$\trip{\mred{\msqu}}{\mgree{\msqu}}{\mblue{\msqu}} \trip{\mred{\mdia}}{\mgree{\mdia}}{\mblue{\mdia}}$\,$ & $\, x_2 \Ph \,$ & \multicolumn{4}{c@{\vrule width.4pt}}{$0$} & $\, 1 \,$ & \multicolumn{3}{c@{\vrule width1pt}}{$\pm 1 \;\;$} & $\, 1 \,$ & $q_{II}$ & $\, 6 \,$ \\
\noalign{\hrule height .2pt}
$\,$\trip{\mred{\ssqu}}{\mgree{\ssqu}}{\mblue{\ssqu}} \trip{\mred{\sdia}}{\mgree{\sdia}}{\mblue{\sdia}}$\,$ & $\, x_3 \Ph \,$ & \multicolumn{4}{c@{\vrule width.4pt}}{$0$} & $\, 0 \,$ & \multicolumn{3}{c@{\vrule width1pt}}{$\pm (1 \;\;\; 1) \;\;$} & $\, 1 \,$ & $q_{III}$ & $\, 6 \,$ \\
\noalign{\hrule height 1.0pt}
\end{tabular}
\caption{The $240$ roots of $E8$ assigned elementary particle labels according to $F4$ and $G2$ subgroups.\label{e8tmedium}}
}

\noindent
The $E8$ root system was first described as a polytope by Thorold Gosset in 1900,\cite{Goss} and the triacontagonal projection plotted by hand in 1964. This plot,\cite{Rich} now with elementary particle symbols assigned to their associated roots according to Table \ref{e8tmedium}, is shown in Figure \ref{e8gf}, with lines drawn between triality partners.

\pagebreak
$\vp{|}$
\FIGURE{
\epsfig{file=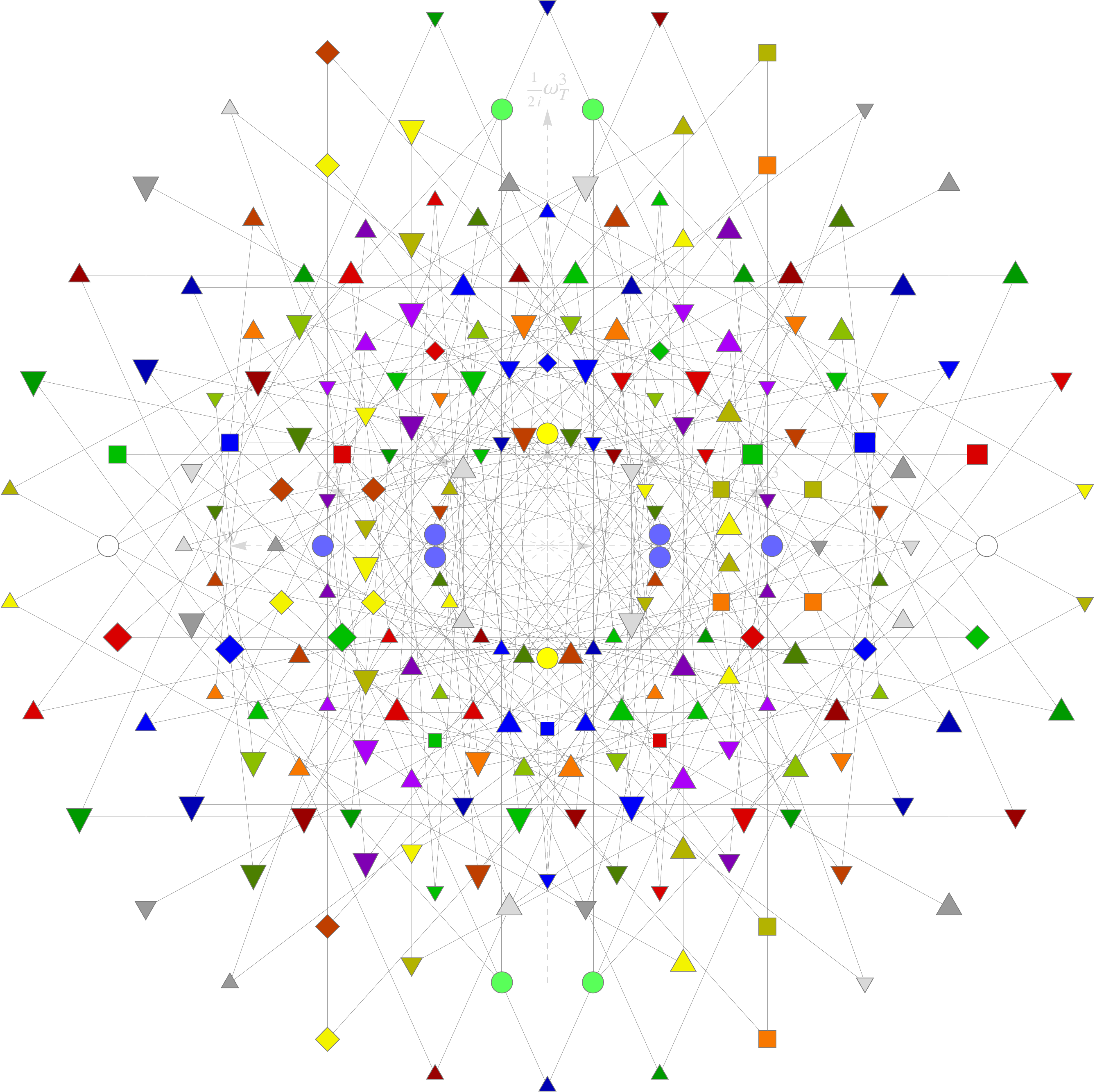, width=6in}
\caption{The $E8$ root system, with each root assigned to an elementary particle field.\label{e8gf}}
}
\pagebreak

The interactions between all standard model and gravitational fields correspond to the Lie brackets between elements of the $E8$ Lie algebra, and thus to the addition of $E8$ roots. The Lie algebra breaks into the standard model (\ref{smalgebra}) as
\begin{eqnarray*}
e8 &=& f4 + g2 + 26 \times 7 \\
&=& \lp so(7,1) + (8+8+8) \rp + \lp su(3) + 3 + \bar{3} \rp 
+ \lp 8+8+8+1+1 \rp \times \lp 3 + \bar{3} + 1 \rp \\
&=& so(7,1) + (su(3) + u(1)) + (8+8+8) \times (3+\bar{3}+1+\bar{1}) + u(1) + 3 \times (3+\bar{3})
\end{eqnarray*}
The $26$ is the the traceless exceptional Jordan algebra --- the smallest irreducible representation space of $F4$ --- and the $7$ is the smallest irreducible representation space of $G2$. Each $8$ is the $8_{S+}$, $8_{S-}$, or $8_V$ of $so(7,1)=d4$. And the $3$ and $\bar{3}$ are in the defining representation space of $su(3)=a2$. The last two terms in the last line above represent new particle fields not in the standard model,
$$
\f{w} \; \in \; \f{u}(1) \qquad \f{x} \Ph \; \in \; 3 \times (3+\bar{3})
$$
The new $\f{x} \Ph$ field carries weak hypercharge and color, has three generations, and couples leptons to quarks. 

This breakdown of $E8$ is possible because $F4$ is the centralizer of $G2$ in $E8$,
$$
F4 = C_{E8}(G2)
$$
To display this subalgebra structure, the $E8$ root system may be rotated in eight dimensions, projected to two, and plotted, as shown in Figures \ref{f4tog2f} and \ref{g2tof4f}.\footnote{An animation of this rotation is available at \href{http://deferentialgeometry.org/anim/e8rotation.mov}{http://deferentialgeometry.org/anim/e8rotation.mov}} In these plots, the root coordinates have been transformed by a rotation,
$$
\lb
\begin{array}{c}
\ha \om_L^3 \\
\ha \om_R^3 \\
W^3 \\
B_1^3 \\
w \\
B_2 \\
g^3 \\
g^8
\end{array}
\rb
=
\lb
\begin{array}{cccccccc}
\fr{1}{\sqrt{2}} & \fr{1}{\sqrt{2}} & & & & & & \\
\fr{-1}{\sqrt{2}} & \fr{1}{\sqrt{2}} & & & & & & \\
& & \fr{1}{\sqrt{2}} & \fr{1}{\sqrt{2}} & & & & \\
& & \fr{-1}{\sqrt{2}} & \fr{1}{\sqrt{2}} & & & & \\
& & & & 1 & & & \\
& & & & & \fr{-1}{\sqrt{3}} & \fr{-1}{\sqrt{3}} & \fr{-1}{\sqrt{3}} \\
& & & & & \fr{-1}{\sqrt{2}} & \fr{1}{\sqrt{2}} & 0 \\
& & & & & \fr{-1}{\sqrt{6}} & \fr{-1}{\sqrt{6}} & \fr{\sqrt{2}}{\sqrt{3}}
\end{array}
\rb
\lb
\begin{array}{c}
\fr{1}{2i} \om_T^3 \\
\ha \om_S^3 \\
U^3 \\
V^3 \\
w \\
x \\
y \\
z
\end{array}
\rb
$$
equivalent to the redefinition of the Cartan subalgebra generators according to (\ref{su3rot}) and (\ref{rot2}). Since the spaces containing the $F4$ and $G2$ root systems are orthogonal in $E8$, these plots of $E8$ showing a rotation between the two are especially pretty and convenient for identifying interactions between particles. Also, the central cluster of $72$ roots in Figure \ref{g2tof4f} is the $E6$ root system, which acts on each of the three colored and anti-colored $27$ element clusters of the exceptional Jordan algebra.

\pagebreak
$\vp{|}$
\FIGURE{
\epsfig{file=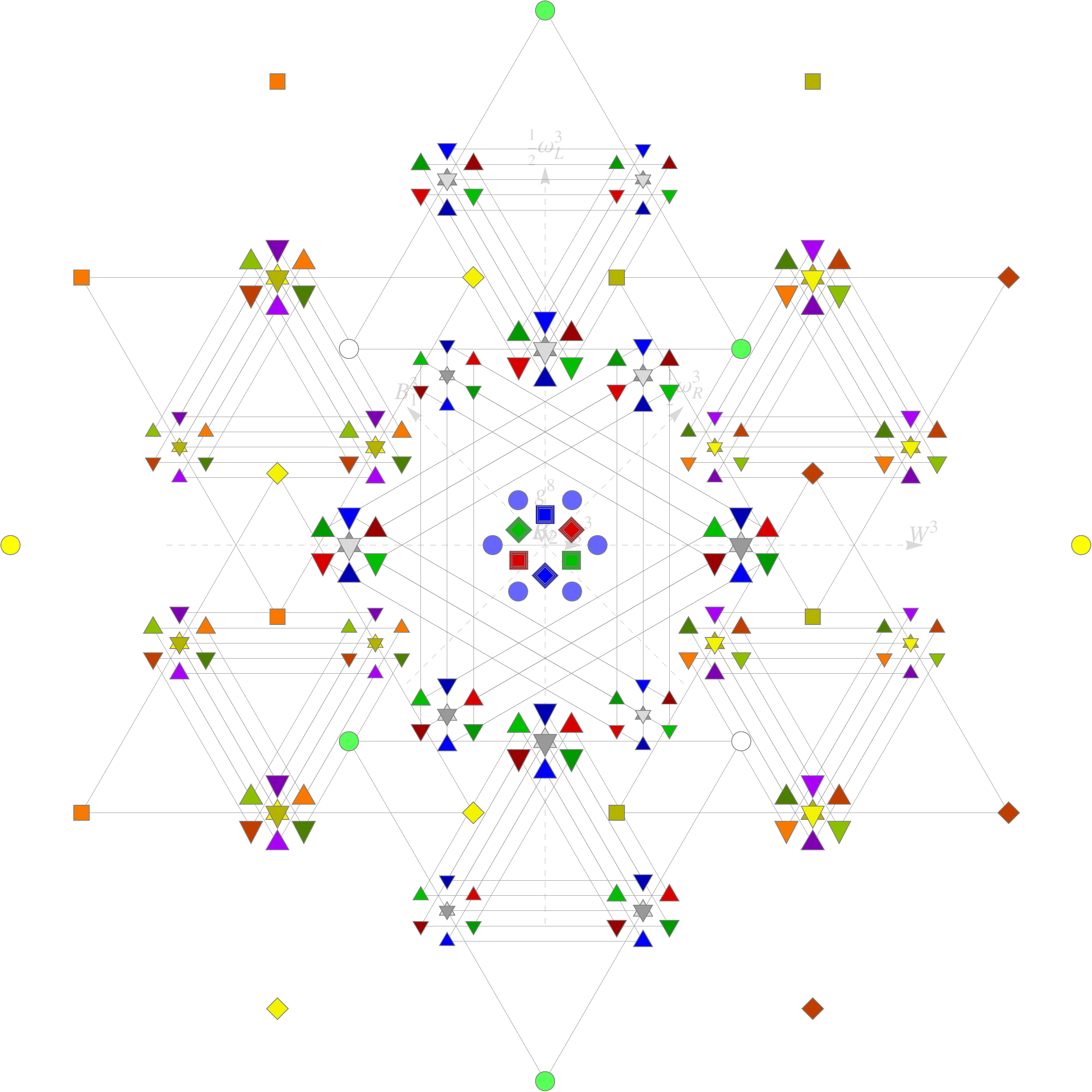, width=6in}
\caption{The $E8$ root system, rotated a little from $F4$ towards $G2$.\label{f4tog2f}}
}
\pagebreak

$\vp{|}$
\FIGURE{
\epsfig{file=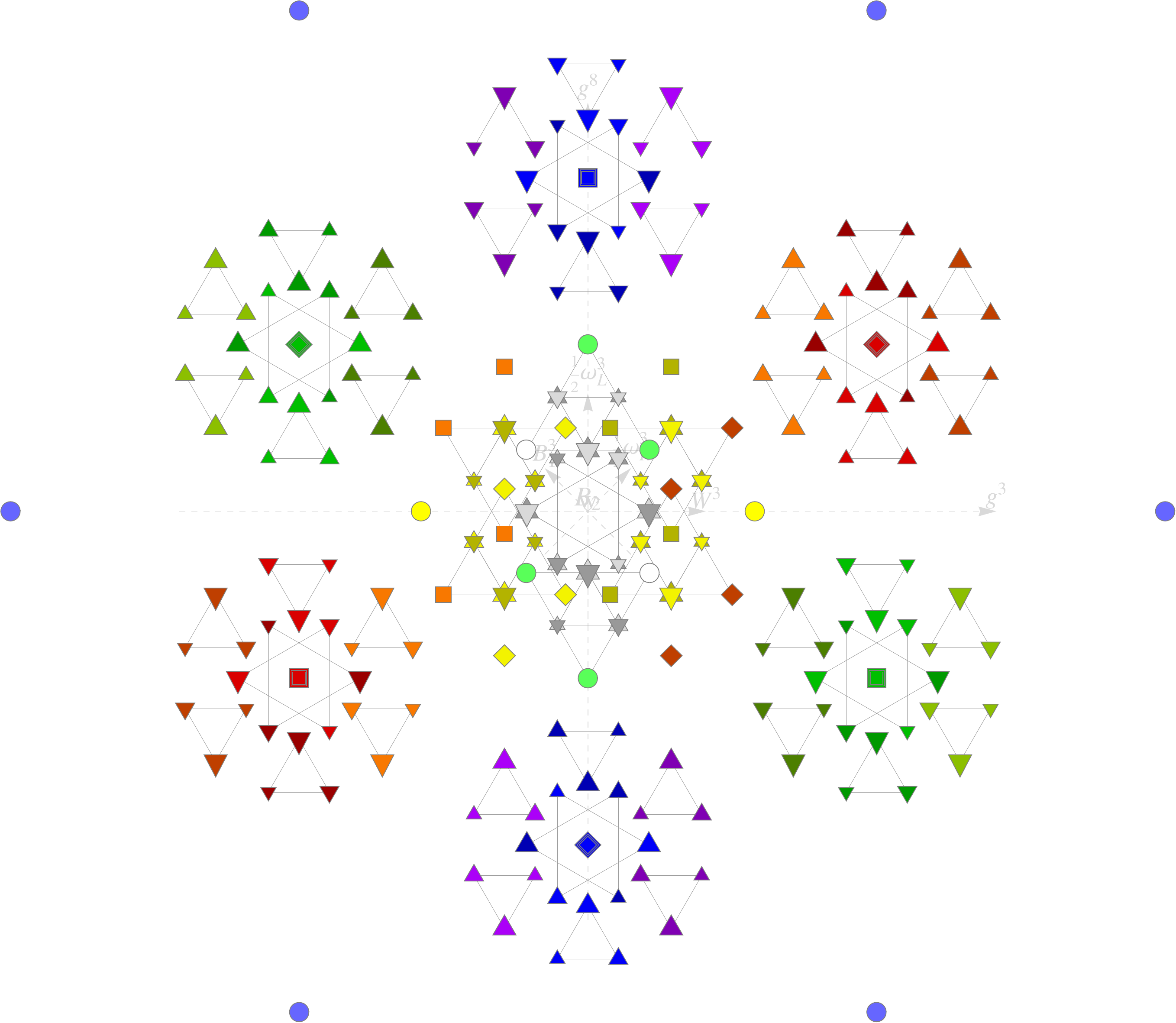, width=6in}
\caption{The $E8$ root system, rotated a little from $G2$ towards $F4$, showing $E6$.\label{g2tof4f}}
}
\pagebreak

Inspecting the $e6$ subalgebra of $e8$ reveals how the fermions and anti-fermions --- up to now described as living in real representations --- are combined in complex representations.\cite{Gurs} The $e6$ decomposes to graviweak $so(7,1)$ acting on three complex generations of fermions as
\begin{eqnarray*}
e6 &=& f4 + (8+8+8) \times \bar{1} + u(1) + u(1) \\
&=& so(7,1) + (8+8+8) \times (1+\bar{1}) + u(1) + u(1) \\
&=& so(9,1) + u(1) + 16_{S\mathbb{C}}
\end{eqnarray*}
in which the final $u(1)$ is the complex structure, $i$, related to the $w$ quantum number, and the $16_{S\mathbb{C}}$ is a complex spinor acted on by the $so(9,1)$.

Although considering its $e6$, $f4$, and $g2$ subalgebras is useful, the $E8$ Lie algebra may be broken down to the standard model via a more direct route,\cite{Baez}
\begin{eqnarray*}
e8 &=& so(7,1) + so(8) + ( 8_{S^+} \!\times 8_{S^+} ) + ( 8_{S^-} \!\times 8_{S^-} ) + ( 8_V \!\times 8_V ) \\
&=& so(7,1) + \big( su(3) + u(1) + u(1) + 3 \times (3+\bar{3}) \big) + (8+8+8)\times(3+\bar{3}+1+\bar{1}) 
\end{eqnarray*}
\FIGURE[r]{
\epsfig{file=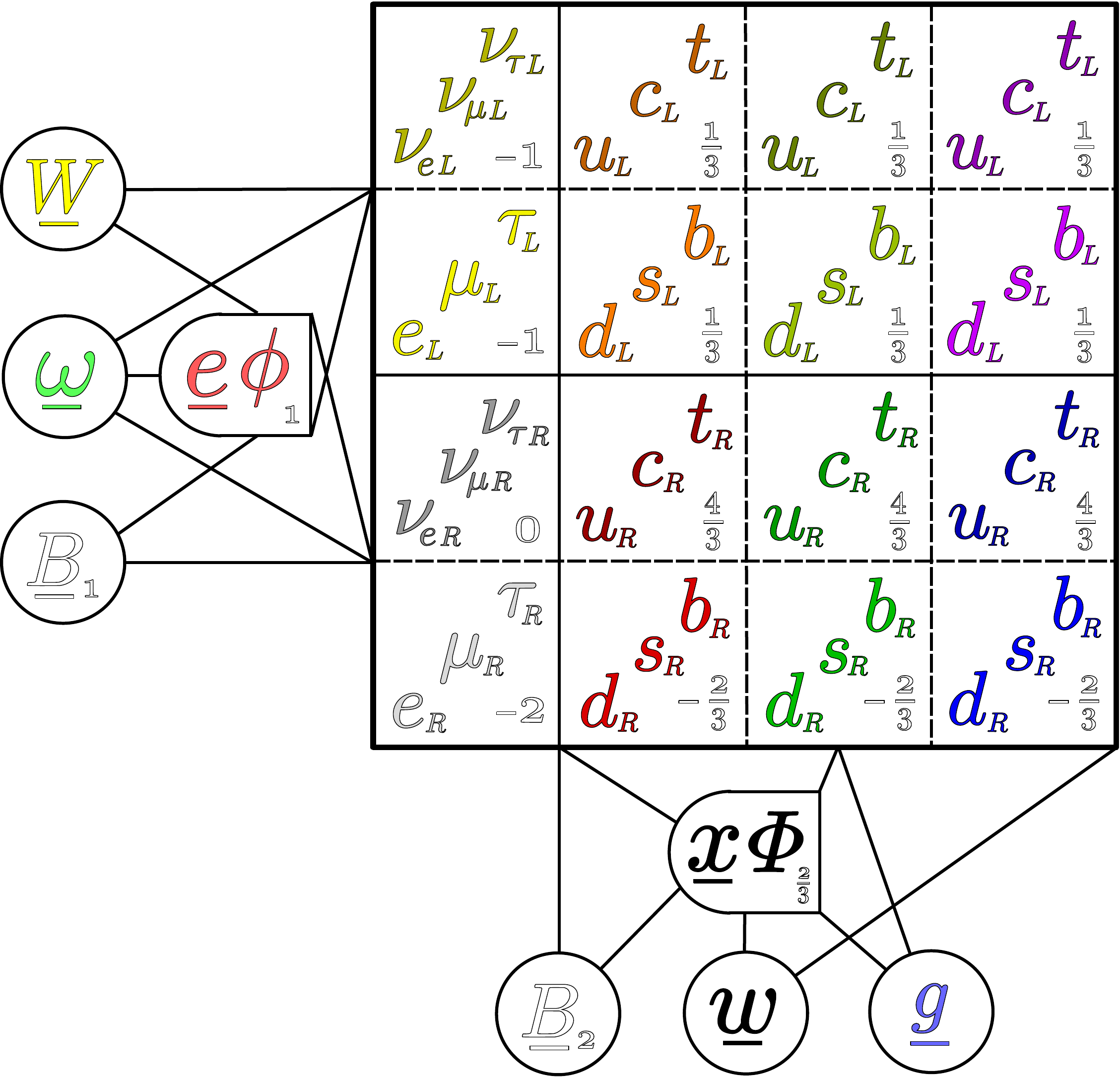, width=3.36in} $\!\!\!$
\caption{A periodic table of $E8$.\label{e8pt}}
}
\noindent
This decomposition is directly visible in Table \ref{e8tmedium}, in which the first four coordinate axes are of $so(7,1)$ and the last four are of $so(8)$. The $so(7,1)$ decomposes into the graviweak fields, and the $so(8)$ decomposes into strong $su(3)$, $u(1)_{B-L}$, and new fields via the embedding of $su(4)$ in $so(8)$. A matched triality rotation of $so(7,1)$ and $so(8)$ relates the three generations of fermions.

The Lie algebra structure of $E8$, and its relation to the structure of the standard model, is depicted in Figure \ref{e8pt} --- a periodic table of $E8$. A comparison of this structure with Figure \ref{ptsm} shows the extremely close fit to the standard model, with only a handful of new particles suggested by the structure of $E8$.

\subsubsection{New particles}

After all algebraic elements of the standard model have been fit to the $E8$ Lie algebra there are a few $e8$ elements remaining, representing new, non-standard particles. There are two new quantum numbers, $X$ and $w$, representing the Pati-Salam partner to weak hypercharge and a new quantum number related to generations. Each of these corresponds to new $u(1)$ valued fields, $\f{X}$ and $\f{w}$, which presumably have large masses impeding their measurement. The use of the Pati-Salam model also implies a non-standard pair of fields, $\f{B}{}_1^\pm$, interacting with right-chiral fermions. In addition, there is a new field, $\f{x}\Ph$, interacting with leptons and quarks. This field factors into three generations, $\f{x}{}_{1/2/3}$, corresponding to different $w$ quantum numbers, and a new Higgs scalar, $\Ph$, for each color and anti-color. The new field, $\f{x}\Ph$, is a joining of $\f{x}$ and $\Ph$ in the same way $\f{e}\ph$ is a joining of the gravitational frame, $\f{e}$, and the Higgs, $\ph$.

Since the frame-Higgs is a composite field --- a simple bivector --- its degrees of freedom do not exhaust the algebraic sector it inhabits. Specifically, $\f{e}\ph = \f{e}^\mu \ph^\nu \Ga_\mu \Ga'_\nu$ uses $16$ algebraic elements but, because it is simple, has only $4$ (for $\f{e}=\f{e}^\mu \Ga_\mu$) plus $4$ (for $\ph=\ph^\nu \Ga'_\nu$) equals $8$ algebraic field degrees of freedom. How or why these $16$ algebraic elements are restricted is not understood --- but this restriction is necessary to recover the standard model and gravity. Because the $18$ algebraic degrees of freedom inhabited by $\f{x}\Ph$ appear amenable to the same sort of factorization as $\f{e}\ph$ (see Table \ref{e8tmedium}), it is natural to factor it into three $\f{x}$ fields and three colored and three anti-colored Higgs fields, $\Ph$. It could be possible that this new $\f{x}\Ph$ gives different masses to the different generations of quarks and leptons, producing the CKM and PMNS matrices. Also, since it mixes leptons and quarks, the existence of this field predicts proton decay, as does any grand unified theory.

The interactions between the new fields, $\f{w}$ and $\f{x}\Ph$, are analogous to the interactions between the gravitational spin connection and the frame-Higgs, $\f{\om}$ and $\f{e}\ph$. This suggests that a better understanding of the triality relationship between generations will involve how these two sets of fields may be more intimately related.

\subsubsection{$E8$ triality}

The specific triality matrix chosen to rotate between the fermion generations, in the coordinates $\{ \ha \om_L^3, \ha \om_R^3, W^3, B_1^3, w, B_2, g^3, g^8 \}$, is
$$
{\scriptsize T =
\lb
\begin{array}{cccccccc}
0 & 0 & 0 & 1 &  &  &  &  \\
1 & 0 & 0 & 0 &  &  &  &  \\
0 & 0 & 1 & 0 &  &  &  &  \\
0 & 1 & 0 & 0 &  &  &  &  \\
 &  &  &  & \fr{-1}{\sqrt{2}} & \fr{-\sqrt{3}}{2} &  &  \\
 &  &  &  & \fr{\sqrt{3}}{2} & \fr{-1}{2} &  &  \\
 &  &  &  &  &  & 1 & 0 \\
 &  &  &  &  &  & 0 & 1
\end{array}
\rb }
$$
This is a somewhat arbitrary choice, selected for leaving $W^3$ and color invariant. Once the first generation of fermions, with correct charges and spins, are assigned to elements of $e8$, this $T$ rotates them to the second and third generations. The second and third generations only have the correct spins and charges when considered as equivalent under this $T$. When considered as independent fields with $E8$ quantum numbers, irrespective of this triality relationship, the second and third generation of fields do not have correct charges and spins. The $W^3$ and color charges are invariant under our choice of $T$ but the spins and hypercharges are only correct through triality equivalence. This relationship between fermion generations and triality is the least understood aspect of this theory.

It is conceivable that there is a more complicated way of assigning three generations of fermions to the $E8$ roots to get standard model quantum numbers for all three generations without triality equivalence. There is such an assignment known to the author that gives the correct hypercharges for all three generations, but it is not a triality rotation and it produces unusual spins. A correct description of the relationship between triality and generations, if it exists, awaits a better understanding.

\section{Dynamics}

The dynamics of a connection is specified by the action functional, $S[\udf{A}]$. Classically, extremizing this action, constrained by boundary data, determines the value of the connection, $\udf{A}(x)$, over a region of the base manifold. The value of the connection may also be used to infer topological properties of the base manifold. Quantum mechanically, the action of a connection over the base manifold determines the probability of experiencing that connection.\cite{Lisi2} Since quantum mechanics is fundamental to our universe, it may be more direct to describe a set of quantum connections as a spin foam, with states described as a spin network. Under more conventional circumstances, the extensive methods of quantum field theory for a non-abelian gauge field may be employed, with propagators and interactions determined by the action. In any case, the dynamics depends on the action, and the action depends on the curvature of the connection.

\subsection{Curvature}

The connection with everything, an $e8$ valued collection of 1-forms and Grassmann fields,
\begin{equation}
\label{e8con}
\udf{A} = \f{H}{}_1 + \f{H}{}_2 + \ud{\Ps}{}_{I} + \ud{\Ps}{}_{II} + \ud{\Ps}{}_{III} \;\; \in \;\; \udf{e8}
\end{equation}
may be broken up into parts matching the standard model,
$$
\begin{array}{rclcl}
\f{H}{}_1 &=& \ha \f{\om} + \fr{1}{4} \f{e}\ph + \f{w}{}_{ew} & \in & \f{so}(7,1) \\
&& \f{\om} & & \; \in \f{so}(3,1) \\
&& \f{e} \ph = (\f{e}{}_1+\f{e}{}_2+\f{e}{}_3+\f{e}{}_4)\times(\ph_{+/0}+\ph_{-/1}) & & \; \in \f{4} \times (2+\bar{2}) \\
&& \f{w}{}_{ew} = \f{W} + \f{B}{}_1 & & \; \in \f{su}(2)_L + \f{su}(2)_R \\[.5em]
\f{H}{}_2 &=& \f{w} + \f{B}{}_2 + \f{x} \Ph + \f{g} & \in & \f{so}(8) \\
&& \f{w} + \f{B}{}_2 & & \; \in \f{u}(1) + \f{u}(1)_{B-L} \\
&& \f{x} \Ph = (\f{x}{}_{1}+\f{x}{}_{2}+\f{x}{}_{3})\times(\Ph^{r/g/b} + \Ph{}^{\bar{r}/\bar{g}/\bar{b}}) & & \; \in \f{3} \times (3+\bar{3}) \\
&& \f{g} & & \; \in \f{su}(3) \\[.5em]
\ud{\Psi}{}_{I} &=& \ud{\nu}{}_e + \ud{e} + \ud{u} + \ud{d} & \in & 8_{S+} \!\times 8_{S+} \\
\ud{\Psi}{}_{II} &=& \ud{\nu}{}_\mu + \ud{\mu} + \ud{c} + \ud{s} & \in & 8_{S-} \!\times 8_{S-} \\ 
\ud{\Psi}{}_{III} &=& \ud{\nu}{}_\ta + \ud{\ta} + \ud{t} + \ud{b} & \in & 8_{V} \times 8_{V} \\
\end{array}
$$
The curvature of this connection, an $e8$ valued collection of 2-forms and Grassmann 1-forms,
\begin{equation}
\label{curvature}
\begin{array}{rcl}
\udff{F} &=& \f{d} \udf{A} + \ha [ \udf{A} , \udf{A} ] \\
&=& \f{d} \udf{A} + \udf{A} \udf{A} \\
&=& \ff{F}{}^1+\ff{F}{}^2+ \f{D} \ud{\Ps}{}_{I} + \f{D} \ud{\Ps}{}_{II} + \f{D} \ud{\Ps}{}_{III}
\end{array}
\end{equation}
may be computed and broken up into standard model parts. The $so(7,1)$ part of the curvature,
$$
\ff{F}{}^1 = \ff{F}{}^G + \ff{F}{}^{gw} + \ff{F}{}^{ew}
$$
includes the gravitational $so(3,1)$ part, the mixed graviweak $4 \times (2+\bar{2})$ part, and the electroweak $su(2)_L+su(2)_R$ part. The gravitational $so(3,1)$ part of the curvature is
\begin{equation}
\label{FG}
\ff{F}{}^G = \ha \big( (\f{d} \f{\om} + \ha \f{\om} \f{\om}) + \fr{1}{8} \f{e} \ph \f{e} \ph \big)
= \ha \big( \ff{R} - \fr{1}{8} \f{e} \f{e} \ph^2 \big)
\end{equation}
in which $\ff{R}$ is the Riemann curvature 2-form, $\f{e} \f{e}$ is the spacetime area bivector, and $\ph^2$ is the amplitude of the Higgs squared. The mixed graviweak $4 \times (2+\bar{2})$ part is
\begin{equation}
\label{Fgw}
\ff{F}{}^{gw} =  (\f{d} \f{e} + \ha [\f{\om},\f{e}]) \ph - \f{e} (\f{d} \ph + [\f{W}+\f{B}{}_1,\ph]) = \ff{T}\ph - \f{e} \f{D} \ph
\end{equation}
in which $\ff{T}$ is the gravitational torsion and $\f{D}$ is the covariant derivative. And the electroweak $su(2)_L+su(2)_R$ part of the curvature is
\begin{equation}
\label{Few}
\ff{F}{}^{ew} =  (\f{d} \f{W} + \f{W} \f{W}) + (\f{d} \f{B}{}_1 + \f{B}{}_1 \f{B}{}_1)
= \ff{F}{}^W + \ff{F}{}^{B_1}
\end{equation}
The $so(8)$ part of the curvature,
\begin{equation}
\label{Fso8}
\ff{F}{}^2 = \ff{F}{}^{w} + \ff{F}{}^{B_2} + \ff{F}{}^{x} + \ff{F}{}^{g} + \f{x}\Ph \f{x}\Ph
\end{equation}
includes the $u(1)$ and $u(1)_{B-L}$ parts, the mixed $3\times(3+\bar{3})$ part, and the strong $su(3)$ part. The last term does not easily separate --- $\f{x}\Ph \f{x}\Ph$ contributes to all three parts of $\ff{F}{}^2$. The $u(1)$ and $u(1)_{B-L}$ parts are
$$
\ff{F}{}^{w} = \f{d} \f{w} \qquad  \ff{F}{}^{B_2} = \f{d} \f{B}{}_2 
$$
The mixed $3\times(3+\bar{3})$ part is
$$
\ff{F}{}^{x} = \big( \f{d} \f{x} \!+\! [ \f{w} \!+\! \f{B}{}_2, \f{x} ] \big) \Ph - \f{x} \big( \f{d} \Ph \!+\! [ \f{g}, \Ph ] \big)
= (\f{D} \f{x}) \Ph \!-\! \f{x} \f{D} \Ph
$$
And the strong $su(3)$ part is
$$
\ff{F}{}^{g} = \f{d} \f{g} + \f{g} \f{g}
$$
Due to the exceptional structure of $e8$, the fermionic part of the curvature for the first generation is
\begin{eqnarray*}
\f{D} \ud{\Psi} &=& \f{d} \ud{\Psi} + [ \f{H}{}_1 + \f{H}{}_2  , \ud{\Psi}] \\ 
&=& \big( \f{d} + \ha \f{\om} + \fr{1}{4} \f{e}\ph \big) \ud{\Ps}
+ \f{W} \ud{\Ps}{}_L + \f{B}{}_1 \ud{\Ps}{}_R - \ud{\Ps} \big( \f{w} + \f{B}{}_2 + \f{x} \Ph \big) - \ud{\Ps}{}_q \, \f{g}
\vp{|^{\Big(}}
\end{eqnarray*}
with $\f{D}$ the covariant massive Dirac derivative in curved spacetime. The second and third fermionic generation parts of this curvature are similar.

\subsection{Action} 

The most conservative approach to specifying the dynamics is to write down an action agreeing with the known standard model and gravitational action while satisfying our desire for minimalism. With these two motivations in mind, an action for everything can be economically expressed as a modified BF theory action over a four dimensional base manifold,
\begin{equation}
\label{action}
S = \int \big< \ff{\od{B}} \udff{F} + \fr{\pi G}{4} \ff{B}{}^G \ff{B}{}^G \ga - \ff{B'} \ff{*B'} \big>
\end{equation}
in which $\udff{F}$ is the curvature (\ref{curvature}), $\ff{\od{B}} = \ff{B} + \fff{\od{B}}$ is an $e8$ valued collection of 2-form and anti-Grassmann 3-form Lagrange multiplier fields, $\ff{B}{}^G$ is the $so(3,1)$ part of $\ff{B}$, $\ff{B'}$ is the rest of $\ff{B}$, $\ga = \Ga_1 \Ga_2 \Ga_3 \Ga_4$ is the Clifford algebra volume element, $*$ is the Hodge star, and $<>$ takes the scalar part (the trace).

After varying $\ff{B}$ and plugging it back in (\ref{action}), this action --- up to a boundary term --- is
\begin{equation}
\label{action2}
S = \int \big< \fff{\od{B}} \f{D} \ud{\Ps}
+ \nf{e} \fr{1}{16 \pi G} \ph^2  \big( R - \fr{3}{2} \ph^2 \big) - \fr{1}{4} \ff{F'} \ff{*F'} \big>
\end{equation}
in which $\nf{e}$ is the spacetime volume 4-form, $R$ is the gravitational scalar curvature, and $\ff{F'}$ is the non-$so(3,1)$ part of $\ff{F}_1$ and $\ff{F}_2$. This is recognizable as the action for the standard model and gravity, with a cosmological constant related to the Higgs vacuum expectation value,
$$
\La = \fr{3}{4} \ph^2
$$
The details of the action, and its agreement with the standard model and general relativity, can be worked out for each sector of the $E8$ Lie algebra.

\subsubsection{Gravity}
The modified BF action for gravity was discovered by MacDowell and Mansouri in 1977, \cite{MacD} and revived by Smolin, Starodubtsev, and Freidel during their work on loop quantum gravity.\cite{Smol, Frei} The remarkable and surprising fact that gravity, described by the spin connection, $\f{\om}$, and frame, $\f{e}$, can be described purely in terms of a unified connection, $\f{\om} + \f{e}$, was the seed idea that led to the unification of all fields in a single connection.\cite{Lisi}

The gravitational part of the action (\ref{action}) is
$$
S_G = \int \big< \ff{B}{}^G \ff{F}{}^G + \fr{\pi G}{4} \ff{B}{}^G \ff{B}{}^G \ga \big>
$$
in which the gravitational part of the curvature (\ref{FG}) is
$$
\ff{F}{}^G = \ha \big( \ff{R} - \fr{1}{8} \f{e} \f{e} \ph^2 \big)
\;\; \in \;\; \ff{so}(3,1)
$$
Extremizing the action under variation of the gravitational part of the Lagrange multiplier, $\de \ff{B}{}^G$, requires
$$
\ff{B}{}^G = \fr{2}{\pi G} \ff{F}{}^G  \ga = \fr{1}{\pi G} \big( \ff{R} - \fr{1}{8} \f{e}\f{e} \ph^2 \big) \ga
$$
and plugging this back into the action gives
$$
S_G = \fr{1}{\pi G} \int \big< \ff{F}{}^G \ff{F}{}^G \ga \big>
=
\fr{1}{4 \pi G} \int \big< \big( \ff{R} - \fr{1}{8} \f{e}\f{e} \ph^2 \big) \big( \ff{R} - \fr{1}{8} \f{e}\f{e} \ph^2 \big) \ga \big>
$$
Multiplying this out gives three terms. The term quadratic in the Riemann curvature is the Chern-Simons boundary term,
$$
\big< \ff{R} \ff{R} \ga \big> = \f{d} \big< \big( \f{\om} \f{d} \f{\om} + \fr{1}{3} \f{\om} \f{\om} \f{\om} \big) \ga \big> 
$$
Dropping this, the other two terms give the Palatini action for gravity,
\begin{eqnarray*}
S_G &=& \fr{1}{16 \pi G} \int \Big\{ \fr{1}{12} \big< \f{e}\f{e} \f{e}\f{e} \ga \big> \ph^4 - \big< \ff{R} \f{e}\f{e} \ga \big> \ph^2 \Big\}  \\
&=&
\fr{1}{16\pi G} \int \nf{e} \, \ph^2 \lp R - \fr{3}{2} \ph^2 \rp
\end{eqnarray*}
equal to the Einstein-Hilbert action with cosmological constant, $\La=\fr{3}{4}\ph^2$. The magnitude of the Higgs, $\sqrt{\ph^2}$, is a conformal factor that can be absorbed into the magnitude of the frame. The vacuum solution to Einstein's equation with positive cosmological constant is de Sitter spacetime  ($\ff{R} = \fr{\La}{6} \f{e} \f{e}$ and $R=4 \La$), which should be considered the background vacuum spacetime for particle interactions in this theory. Since the symmetry of this spacetime is $so(4,1)$ and not the Poincar\'{e} group, the Coleman-Mandula theorem does not apply to restrict the unification of gravity within the larger group.

It should be emphasized that the connection (\ref{e8con}) comprises all fields over the four dimensional base manifold. There are no other fields required to match the fields of the standard model and gravity. The gravitational metric and connection have been supplanted by the frame and spin connection parts of $\udf{A}$. The Riemannian geometry of general relativity has been subsumed by principal bundle geometry --- a significant mathematical unification. Devotees of geometry should not despair at this development, as principal bundle geometry is even more natural than Riemannian geometry. A principal bundle with connection can be described purely in terms of a mapping between tangent vector fields (diffeomorphisms) on a manifold, without the ab initio introduction of a metric.

\subsubsection{Other bosons}

The part of the action (\ref{action2}) for the bosonic, non-$so(3,1)$ parts of the connection is
$$
S' = - \int \fr{1}{4} \big< \ff{F'} \ff{*F'} \big> = S_{gw} + S_{ew} + S_2
$$
in which the relevant parts of the curvature (\ref{curvature}) are the mixed graviweak part (\ref{Fgw}), the electroweak part (\ref{Few}), and the $so(8)$ part (\ref{Fso8}). The mixed graviweak part of the action is
\begin{eqnarray*}
S_{gw} &=& - \int \fr{1}{4} \big< \ff{F}{}^{gw} \ff{*F}{}^{gw} \big> \\
&=& - \int \fr{1}{4} \big< (\ff{T}\ph-\f{e}\f{D}\ph) * (\ff{T}\ph-\f{e}\f{D}\ph) \big> \\
&=& \int \Big\{
\fr{1}{4} \big< \ff{T} * \ff{T} \big> \ph^2 
+ \big< \ph (\f{D} \ph) \f{e} * \ff{T} \big>
+ \fr{3}{4} \big< (\f{D} \ph) * (\f{D} \ph) \big>
 \Big\} \\
\end{eqnarray*}
which includes the kinetic Higgs term and gravitational torsion. The electroweak part of the action is
$$
S_{ew} = - \int \fr{1}{4} \big< \ff{F}{}^{ew} \ff{*F}{}^{ew} \big>
= - \int \fr{1}{4} \big< \ff{F}{}^{W} \ff{*F}{}^{W} \big>
- \int \fr{1}{4} \big< \ff{F}{}^{B_1} \ff{*F}{}^{B_1} \big>
$$
And the $so(8)$ part of the action is
\begin{eqnarray*}
S_2 &=& - \int \fr{1}{4} \big< \ff{F}{}^2 \ff{*F}{}^2 \big> \\
&=& - \int \fr{1}{4} \big< \ff{F}{}^{w} \ff{*F}{}^{w} \big>
- \int \fr{1}{4} \big< \ff{F}{}^{B_2} \ff{*F}{}^{B_2} \big>
- \int \fr{1}{4} \big< \ff{F}{}^x \ff{*F}{}^x \big> 
- \int \fr{1}{4} \big< \ff{F}{}^g \ff{*F}{}^g \big> -\\
&&- \int \fr{1}{2} \big<
\big( \ff{F}{}^w + \ff{F}{}^{B_2} + \ff{F}{}^x + \ff{F}{}^g + \f{x}\Ph\f{x}\Ph  \big)
*
\f{x}\Ph\f{x}\Ph
\big>
\end{eqnarray*}
which includes the action for the gluons and a first guess at the action for the new fields. This action for the new fields is speculative at this stage and likely to change as our understanding of their role improves.

The use of the Hodge dual in this part of the action is required for general covariance but seems somewhat awkward from the viewpoint of this $E8$ theory. The Hodge star operator requires the frame part, $\f{e}$, to be extracted from the $E8$ connection, inverted to obtain the coframe, $\ve{e}$, and contracted with the curvature. It would be better if there was a natural justification for this procedure, beyond the necessity to agree with known theory. An improved understanding will likely lead to a modification of this part of the action.

\subsubsection{Fermions}

Choosing the anti-Grassmann Lagrange multiplier 3-form to be $\fff{\od{B}} = \nf{e} \od{\Ps} \ve{e} \,$ in the fermionic part of the action (\ref{action2}) gives the massive Dirac action in curved spacetime,
\begin{eqnarray*}
S_f &=& \int \big< \fff{\od{B}} \f{D} \ud{\Ps} \big> \\
&=& \int \big< \nf{e} \od{\Ps} \ve{e} \big( \f{d} \ud{\Ps} + \f{H}{}_1 \ud{\Ps} - \ud{\Ps} \f{H}{}_2 \big) \big> \\
&=& \int \big< \nf{e} \od{\Ps} \ve{e} \big( ( \f{d} + \ha \f{\om} + \fr{1}{4} \f{e}\ph + \f{W} + \f{B}{}_1 ) \ud{\Ps}
- \ud{\Ps} ( \f{w} + \f{B}{}_2 + \f{x} \Ph + \f{g} )  \big) \big> \\
&=& \int \nf{d^4 x} \, |e| \, \big< \od{\Ps} \ga^\mu (e_\mu)^i \big( \pa_i \ud{\Ps} + \fr{1}{4} \om_i^{\p{i} \mu \nu} \ga_{\mu \nu} \ud{\Ps} + W_i \ud{\Ps} + B_{1i} \ud{\Ps} + \\
&& \hphantom{\int \nf{d^4 x} \, |e| \, \big< \od{\Ps} \ga^\mu (e_\mu)^i \big(} + \ud{\Ps} w_i + \ud{\Ps} B_{2i} + \ud{\Ps} x_i \Ph + \ud{\Ps} g_i \big) + \od{\Ps} \, \ph \, \ud{\Ps} \big>
\end{eqnarray*}
The coframe, $\ve{e}$, in this action contracts with the frame part of the graviweak connection,
$$
\ve{e} \f{e} = \ga^\mu (e_\mu)^i \ve{\pa_i} \f{dx^j} (e_j)^\nu  \ga_\nu
= \ga^\mu (e_\mu)^i (e_i)^\nu  \ga_\nu
= \ga^\mu  \ga_\mu
= 4
$$
to give the standard Higgs coupling term, $\od{\Ps} \, \ph \, \ud{\Ps}$. The new, non-standard $\od{\Ps} \ga^\mu \ud{\Ps} w_\mu$ and $\od{\Ps} \ga^\mu \ud{\Ps} x_\mu \Ph$ terms are not yet well understood but seem promising for recovering the CKM matrix.

This action works very well for one generation of fermions. The action for the other two generations should be similar, but is related by triality in a way that is not presently understood well enough to write down.

\section{Summary}

The ``E8 theory'' proposed in this work is an exceptionally simple unification of the standard model and gravity. All known fields are parts of an $E8$ principal bundle connection,
$$
\begin{array}{rcl}
\udf{A} &=& ( \ha \f{\om} + \fr{1}{4} \f{e} \ph + \f{W} + \f{B}{}_1 ) + ( \f{B}{}_2 + \f{w} + \f{x} \Ph + \f{g} ) + \vp{{}_\big(} \\
&& + \, \big( \ud{\nu}{}_e + \ud{e} + \ud{u} + \ud{d} \big)
+ \big( \ud{\nu}{}_\mu + \ud{\mu} + \ud{c} + \ud{s} \big)
+ \big( \ud{\nu}{}_\ta + \ud{\ta} + \ud{t} + \ud{b} \big)
\end{array}
$$
in agreement with the Pati-Salam $SU(2)_L \times SU(2)_R \times SU(4)$ grand unified theory, with a handfull of new fields suggested by the structure of $E8$. The interactions are described by the curvature of this connection,
$$
\udff{F} = \f{d} \udf{A} + \ha \big[ \udf{A}, \udf{A} \big]
$$
with particle quantum numbers corresponding to the vertices of the $E8$ polytope in eight dimensions. This structure suggests three fermionic generations related by triality,
$$
T \, e = \mu \qquad T \, \mu = \ta \qquad T \, \ta = e
$$
The action for everything, chosen by hand to be in agreement with the standard model, is concisely expressed as a modified BF theory action, 
$$
S = \int \big< \ff{\od{B}} \udff{F}
+ \fr{\pi}{4} \ff{B}{}_G \ff{B}{}_G \ga - \ff{B'} \ff{*B'} \big>
$$
with gravity included via the MacDowell-Mansouri technique. The theory has no free parameters. The coupling constants are unified at high energy, and the cosmological constant and masses arise from the vacuum expectation values of the various Higgs fields,
$$
g_1 = \sqrt{\nfr{3}{5}} \qquad g_2=1 \qquad g_3=1 \qquad \La=\fr{3}{4}\ph^2 \qquad M \sim \ph_0 , \ph_1, \Ph \dots 
$$
In sum, everything is described by the pure geometry of an $E8$ principal bundle, perhaps the most beautiful structure in mathematics.

\section{Discussion and Conclusion}

There are a remarkable number of ``coincidences'' that work exactly right to allow all known fields to be unified as parts of one connection. The factors of $\ha$ and $\fr{1}{4}$ multiplying the spin connection and frame-Higgs result in the correct expressions for the gravitational Riemann curvature and the covariant Dirac derivative in curved spacetime. The fermions fit together perfectly in chiral representations under graviweak $so(7,1)$, and the frame-Higgs has all the correct interactions. This frame-Higgs naturally gets a $\ph^4$ potential and produces a positive cosmological constant. Finally, and most impressively, the fit of all fields of the standard model and gravity to $E8$ is very tight. The structure of $E8$ determines exactly the spinor multiplet structure of the known fermions.

There are also aspects of this theory that are poorly understood. The relationship between fermion generations and triality is suggested by the structure of $E8$ but is not perfectly clear --- a better description may follow from an improved understanding of the new $\f{w}+\f{x}\Ph$ fields and their relation to $\f{\om}+\f{e}\ph$. This relationship may also shed light on how and why nature has chosen a non-compact form, $E \; IX$, of $E8$. Currently, the symmetry breaking and action for the theory are chosen by hand to match the standard model --- this needs a mathematical justification.

Quantum E8 theory follows the methods of quantum field theory and loop quantum gravity --- though the details await future work. One enticing possibility is that the gravitational and cosmological constants run from large values at an ultraviolet fixed point to the tiny values we encounter at low energies.\cite{Reut,Perc} At the foundational level, a quantum description of the standard model in $E8$ may be compatible with a spin foam description in terms of braided ribbon networks\cite{Bils} through the identification of the corresponding finite groups. And there is a more speculative possibility: if the universe is described by an exceptional mathematical structure, this suggests quantum E8 theory may relate to an exceptional Kac-Moody algebra.\cite{Kac}

The theory proposed in this paper represents a comprehensive unification program, describing all fields of the standard model and gravity as parts of a uniquely beautiful mathematical structure. The principal bundle connection and its curvature describe how the $E8$ manifold twists and turns over spacetime, reproducing all known fields and dynamics through pure geometry. Some aspects of this theory are not yet completely understood, and until they are it should be treated with appropriate skepticism. However, the current match to the standard model and gravity is very good. Future work will either strengthen the correlation to known physics and produce successful predictions for the LHC, or the theory will encounter a fatal contradiction with nature. The lack of extraneous structures and free parameters ensures testable predictions, so it will either succeed or fail spectacularly. If E8 theory is fully successful as a theory of everything, our universe is an exceptionally beautiful shape.

\acknowledgments

The author wishes to thank 
Peter Woit,
Sergei Winitzki,
Lee Smolin,
Tony Smith,
David Richter,
Fabrizio Nesti,
Sabine Hossenfelder,       
Laurent Freidel,
David Finkelstein,
Michael Edwards,
James Bjorken,
Sundance Bilson-Thompson, 
John Baez,
and 
Stephon Alexander
for valuable discussions and encouragement. Some of the work was carried out under the wonderful hospitality of the Perimeter Institute for Theoretical Physics. This research was supported by grant RFP1-06-07 from The Foundational Questions Institute (\href{http://fqxi.org/aw-lisi.html}{fqxi.org}).


\begin{thebibliography}{99}
\bibitem{Tegm} M. Tegmark, ``Is `the theory of everything' merely the ultimate ensemble theory?'' \grqc{9704009}.
\end{thebibliography}
\end{document}